\documentclass[twocolumn]{aastex701}  

\usepackage{float} 
\usepackage[utf8]{inputenc}
\usepackage{verbatim}
\usepackage{graphicx}
\usepackage{natbib}
\usepackage{amsmath}
\usepackage{hyperref}
\usepackage{url}
\usepackage{booktabs}
\usepackage{siunitx}
\usepackage{ragged2e}
\DeclareSIUnit{\parsec}{pc}
\DeclareSIUnit{\kpc}{kpc}
\newcommand{\jt}{\textcolor{blue}}

\newcommand{\numax}{$\nu_{\rm max}$}
\newcommand{\dnu}{$\Delta\nu$}

\usepackage{newtxtext,newtxmath}
\usepackage[italic]{hepnames}

\usepackage{caption}
\usepackage{subcaption}

\begin{document}
\title{Evaluating the Sensitivity of the Age Inferences of Red Giant Stars to Machine Learning Methodology}
\author[0000-0002-4818-7885]{Jamie Tayar}
\affiliation{Department of Astronomy, University of Florida, Gainesville, FL 32611, USA}
\email[show]{jtayar@ufl.edu}

\author[0009-0003-8168-4474]{Carli Mankowski}
\affiliation{Department of Astronomy, New Mexico State University, Las Cruces, NM 88003, USA}
\affiliation{Department of Astronomy, University of Florida, Gainesville, FL 32611, USA}
\email[show]{rose22@nmsu.edu}

\author[0009-0004-5330-3774]{Lara Tunca}
\affiliation{Department of Physics, Oregon State University, Corvallis, OR 97331, USA}
\affiliation{Department of Astronomy, University of Florida, Gainesville, FL 32611, USA}
\email[show]{ltunca@ufl.edu}


\author[0009-0005-8324-0134]{Dante Jordan}
\affiliation{Department of Astronomy, University of Florida, Gainesville, FL 32611, USA}
\email{dsjor06@gmail.com}

\author[0009-0002-2528-2924]{Mia Severino}
\affiliation{Department of Astronomy, University of Florida, Gainesville, FL 32611, USA}
\email{mj.severino@uf.edu}


\author[0009-0001-5961-749X]{Sydney McArthur}
\affiliation{Department of Astronomy, University of Florida, Gainesville, FL 32611, USA}
\email{sydney.mcarthur@ufl.edu}

\author[0000-0001-9416-1072]{Zeina Benton}
\affiliation{Department of Astronomy, University of Florida, Gainesville, FL 32611, USA}
\email{zeinabenton@ufl.edu}

\author[0009-0008-9853-8461]{Sophia Armstrong}
\affiliation{Department of Astronomy, University of Florida, Gainesville, FL 32611, USA}
\email{armstrong.sophia@hotmail.com}

\author[0009-0000-9841-0670]{Alexa Leddy}
\affiliation{Department of Astronomy, University of Florida, Gainesville, FL 32611, USA}
\email{alexaleddy@ufl.edu}

\author[0009-0006-6056-781X]{Emily Bower}
\affiliation{Department of Astronomy, University of Florida, Gainesville, FL 32611, USA}
\affiliation{Physics Department, Lancaster University, Lancaster, LA1 4TB, UK }
\email{e.bower@lancaster.ac.uk}

 \author[0009-0002-3349-5305]{Zabdiel Sanchez}
\affiliation{The Department of Physics and Astronomy, University of Minnesota Duluth, Duluth, MN 55812, USA}
\affiliation{Department of Astronomy, University of Florida, Gainesville, FL 32611, USA}
\email{sanchez.z@ufl.edu}
\email{sanc0638@d.umn.edu}

\author[0009-0006-4155-4802]{Colin Avery}
\affiliation{Department of Astronomy, University of Florida, Gainesville, FL 32611, USA}
\email{colin.avery@ufl.edu}

\author[0009-0004-2037-1279]{Emily Cummings}
\affiliation{Department of Astronomy, University of Florida, Gainesville, FL 32611, USA}
\email{emilycummings1@ufl.edu}

\author[0009-0007-3104-9228]{Joshua Donley}
\affiliation{Department of Astronomy, University of Florida, Gainesville, FL 32611, USA}
\email{jdonley@ufl.edu}

\author[0009-0008-2164-5357]{Rachel Freeman}
\affiliation{Department of Astronomy, University of Florida, Gainesville, FL 32611, USA}
\email{rachel.freeman@ufl.edu}

\author[0009-0004-1831-9906]{David R. Fulcher}
\affiliation{Department of Physics and Astronomy, University of North Carolina at Chapel Hill, Chapel Hill, NC 27599, USA}
\affiliation{Department of Astronomy, University of Florida, Gainesville, FL 32611, USA}
\email{drf@unc.edu}

\author[0009-0005-3791-832X]{Vanessa Hervie}
\affiliation{Department of Astronomy, University of Florida, Gainesville, FL 32611, USA}
\email{vhervie@ufl.edu}

\author[0009-0002-0921-9191]{James Ivey}
\affiliation{Department of Astronomy, University of Florida, Gainesville, FL 32611, USA}
\email{ivey.jw@ufl.edu}

\author[orcid=0009-0009-6602-4703]{Hyde Kenney}
\affiliation{Department of Astronomy, University of Florida, Gainesville, FL 32611, USA}
\email{hkenney@ufl.edu}  

\author[0009-0000-7728-1742]{William MacMillan}
\affiliation{Department of Astronomy, University of Florida, Gainesville, FL 32611, USA}
\email{w.macmillan@ufl.edu}

\author[0009-0000-2445-7644]{Jake Mahoney}
\affiliation{Department of Astronomy, University of Florida, Gainesville, FL 32611, USA}
\email{mahoneyjake@ufl.edu}

\author[0009-0005-8980-8747]{Eve Maramba}
\affiliation{Department of Mechanical and Aerospace Engineering, University of Florida, Gainesville, FL 32611, USA}
\email{emaramba@ufl.edu}

\author[0009-0007-7196-9133]{Erin Philip}
\affiliation{Department of Astronomy, University of Florida, Gainesville, FL 32611, USA}
\email{erinphilip@ufl.edu}

\author[0009-0005-9327-7547]{Yazmeen Simpson}
\affiliation{Department of Astronomy, University of Florida, Gainesville, FL 32611, USA}
\email{yazmeen.simpson@gmail.com}

\author[0009-0007-1375-322X]{Ethan Strojie}
\affiliation{Department of Astronomy, University of Florida, Gainesville, FL 32611, USA}
\email{estrojie@ufl.edu}

\begin{abstract}
Stellar ages are vital for understanding the formation of our galaxy, but they are among the most challenging parameters to measure. Many authors address this by using machine learning models trained on stars of known age. Here we used data for 351,995 stars from Milky Way Mapper Data Release 19 to explore the sensitivity of the inferred ages to 1) neural network hyperparameters, 2) machine learning architecture, and 3) training set. We find that the resulting ages are generally insensitive to the neural network hyperparameters or the machine learning architecture, but are somewhat sensitive to the training set chosen. We also find that ages for the oldest, coolest, and lowest metallicity stars in the sample are most sensitive to the methodology used and the training set chosen. In general, our analysis suggests that even simple neural network models are sufficient for accurate age inference, but future work expanding the available training sets will be an important component of predicting reliable ages for the full galactic population.

\end{abstract}

\section{Introduction}
Currently, much work is being done to try to understand the formation and evolution of our own Milky Way Galaxy. This galactic archaeology \citep{FreemanBlandHawthorn2002} combines information on the positions and kinematics of stars derived from the Gaia Mission \citep{GaiaDR3} with chemical information about their individual element abundances either from Gaia \citep{Andrae2023} or from large ground-based spectroscopic surveys such as {GALactic Archaeology with HERMES} \citep[GALAH, ][]{Buder2021} or the Sloan Digital Sky Survey \citep[SDSS, ][]{Kollmeier2026} to put together a picture of \textit{what} stars are made out of and \textit{where} they are located in the galaxy. This allows the grouping of populations of stars that were born together \citep[e.g. chemical tagging,][]{FreemanBlandHawthorn2002,  Casamiquela2021}, the identification of stars that were accreted from dwarf galaxies that fell into the Milky Way \citep[e.g.][]{Helmi2018, Naidu_2020}, as well as chemical cartography \citep{Hayden2015} that allows us to understand how nucleosynthesis proceeds \citep{Weinberg2019, Griffith2021} and trace chemical enrichment through our galaxy \citep{Andrews2017, JohnsonJ2021}. This allows us to both understand the evolution of our own galaxy, and to use the Milky Way as a tool for exploring the evolution of galaxies in general.

Additional information on \textit{when} the stars were formed allows this chemodynamic picture to be analyzed across cosmic time. We expect star formation and chemical enrichment to proceed differently in different galaxies \citep{Nidever2020, Povick2024}, and we also expect this evolution to proceed differently in different parts of the galaxy, with the bulge enriching faster \citep{ Joyce2023, Miller2025} and the disk forming and enriching from the inside out \citep{Bird2013, Ness2016, Hasselquist2019, FAnders2023, StoneMartinez2025}. 

However, despite their importance, stellar ages cannot generally be measured directly \citep{Soderblom2010}. Star clusters have long been the gold standard of age inference, with efforts on large, well-studied clusters yielding ages to roughly 5 percent accuracy, even when accounting for uncertainties in the underlying physics of the stellar interior \citep{Ying2023,Reyes2024, Ying2025}. However, most stars are not in clusters, so empirical calibrations to infer ages from stellar rotation \citep[gyrochronology, e.g.][]{barnes2007, Curtis2020}, stellar chemistry \citep[e.g. the carbon-to-nitrogen ratio,][]{Martig2016, Ness2016, Spoo2022}, or dynamics \citep{Lu_2021_gyro, Lu2026} have been developed. Stellar evolution models and isochrones \citep[e.g.][]{Bressan2012, Choi2016, Pinsonneault2026} also offer a method to estimate stellar ages that can be calibrated to clusters \citep{Choi2018b} to address uncertainties in stellar interiors \citep{Tayar2022a, Ying2024, Morales2025} and provide high precision results \citep{Godoy-Rivera2021, XiangRix2022}. More recently, asteroseismology, the study of stellar oscillations \citep{KjeldsenBedding1995}, has offered the possibility to precisely estimate large numbers of stellar ages from photometric monitoring \citep[e.g.][]{SilvaAguirre2018,Stokholm2023,Pinsonneault2025}, and calibrating the accuracy of these results in open clusters is an ongoing effort \citep{Corsaro2012, Pinsonneault2018, TayarJoyce2025, Mankowski2026, Howell2026}. 

While there are many different methods to estimate stellar ages, most of them are data-intensive and only work on certain types of stars, making them challenging to apply to the large populations needed for galactic archaeology studies. Many authors have tried to bridge this gap using machine learning, training their results on smaller samples in order to rapidly infer ages for hundreds of thousands or millions of stars. Many such analyses have trained on asteroseismic data for red giants, which can be studied spectroscopically across the Milky Way \citep[e.g.][]{Ness2016, Ciuca2024, LeungBovy2019, FAnders2023, Leung2023}, although others have trained on rotation data \citep{VanLane2025}, chemistry \citep{Hayden2022,Weaver2024, TamamesRodero2025}, or stellar models \citep{garraffo2021,Hon2024,Saunders2024,Boin2026}. Many different machine learning architectures have been explored in this process, with everything from simple neural networks \citep{Tayar2023RNAAS} to more complex Bayesian convolutional neural networks \citep{Mackereth2019}, variational encoder-decoder network \citep{Leung2023}, and normalizing flows \citep{StoneMartinez2025}. This work has generated valuable maps of the chemical and star-formation history of the Milky Way, but questions have been raised about the ability to reliably infer the oldest and youngest ages \citep{Mackereth2019,TingYS2025}, the impact of the training set \citep{Tayar2023RNAAS, Boin2026}, and the agreement between methods \citep{FAnders2023}. 
Given the ubiquity of using machine learning techniques to estimate stellar ages for large samples that can then be used in galactic archaeology to reconstruct the history and evolution of the Milky Way, we think it prudent to explore how the inferred ages depend on the machine learning technique chosen, the hyperparameters used, and the chosen training set in order to provide a more complete understanding of the uncertainties inherent in using such ages.

\setcounter{footnote}{0}
\section{Methods}

In this paper, we are exploring the impact of different machine learning techniques (Section \ref{ssec:methodml}) and different training data sets (Section \ref{ssec:Seismic}) on the ages inferred for an appropriately selected sample (Section \ref{ssec:methodselect}) of the red giants in Milky Way Mapper (Section \ref{ssec:methodmwm}). We validate our results by comparison to open and globular cluster stars (Section \ref{ssec:methodvalidation}). 

\subsection{Milky Way Mapper} \label{ssec:methodmwm}
\begin{figure}[tb]
    \centering  \includegraphics[width=0.47\textwidth, clip=true, trim=0in 0in 0in 0in]{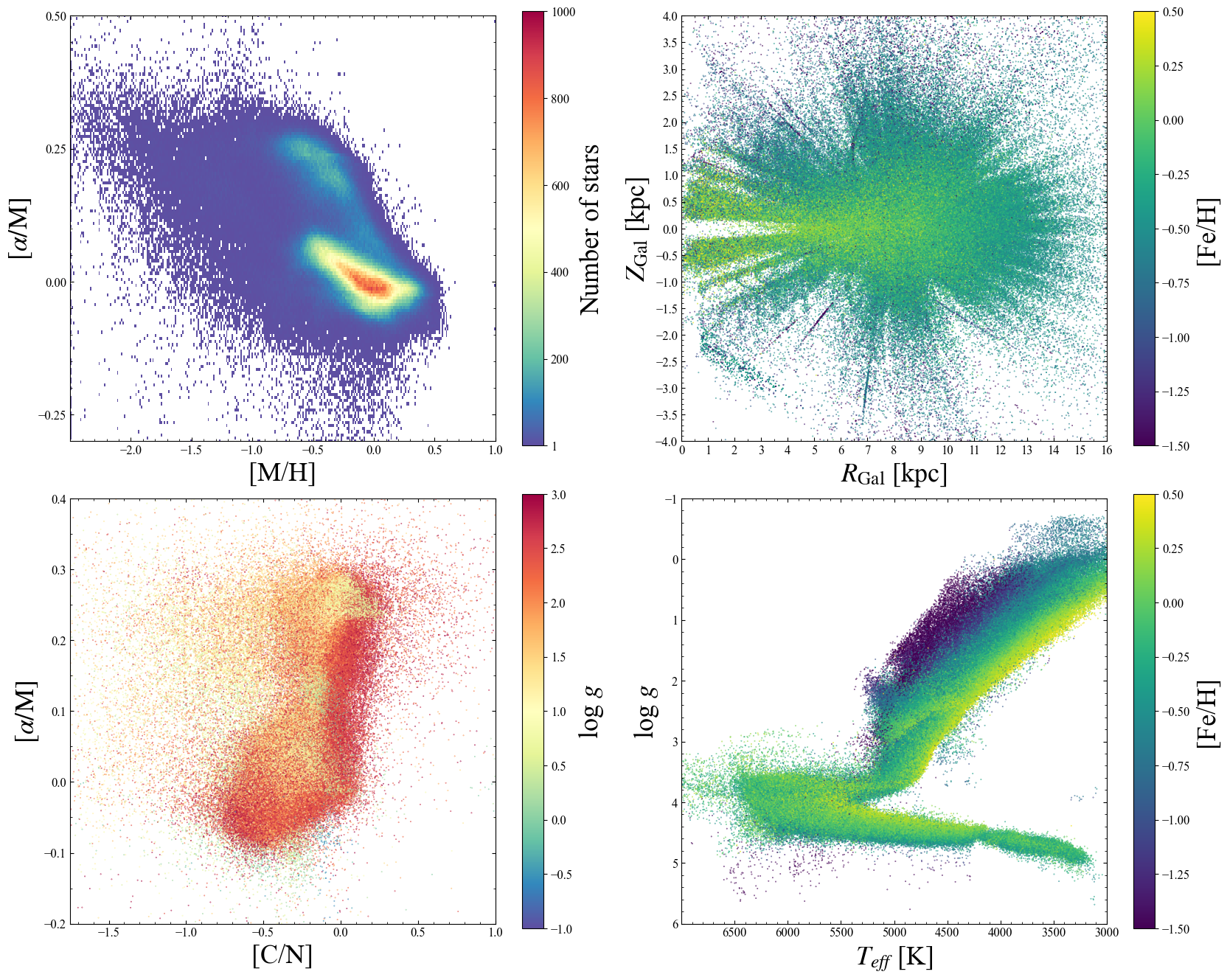}
    \caption{General properties of the MWM DR19 sample. \textbf{Top left:} [$\alpha$/Fe]-[Fe/H] relation, color-coded by the number of stars per bin. 
    \textbf{Top right:} [Fe/H] map in cylindrical coordinates ($\textit{Z}_\text{Gal}$ versus $\textit{R}_\text{Gal}$). \textbf{Bottom left:} [$\alpha$/M]-[C/N] relation, color-coded by log \textit{g}. \textbf{Bottom right:} Kiel diagram (log(g) versus $\textit{T}_\text{eff}$), color-coded by [Fe/H].}
    \label{fig:DR19}
\end{figure}

The majority of our analysis relies on spectroscopic parameters from Data Release 19 \citep{DR19} of the Sloan Digital Sky Survey V \citep{Kollmeier2026}. Specifically, we are using results from the APOGEE \citep{APOGEE_spec_wilson} spectrograph in the H-band, with a resolution of $\sim$22,000 on the SDSS \citep{SDSS_gunn2006} and Dupont telescopes \citep{SDSS_bowen1973}. These results were released as part of the Milky Way Mapper \citep[][J. Johnson et al in prep]{DR18} survey analyzed using the Astra (A. Casey et al. in prep) pipeline run on wavelength-corrected spectra  \citep{Nidever2015, Saydjari2025}. For our analysis, we use the results from the ASPCAP \citep{GarciaPerez2016} pipeline, whose values were then calibrated \citep{Meszaros2025} using the temperature of giants in low extinction fields \citep{Meszaros2025}, the asteroseismic surface gravities from APOKASC-3 \citep{Pinsonneault2025}, and the abundances in star clusters \citep{Otto2026}. This provided us a sample of {963,539} stars, including {703,178} giants with reliable parameters. We show the general properties of the sample in Figure \ref{fig:DR19}, and describe the cuts we made to the sample in Section \ref{ssec:methodselect}.

\subsection{Asteroseismic Ages} \label{ssec:Seismic}
\begin{figure*}[bt]
    \centering  \includegraphics[width=0.9\textwidth,clip=true, trim=0in 0in 0in 0in]{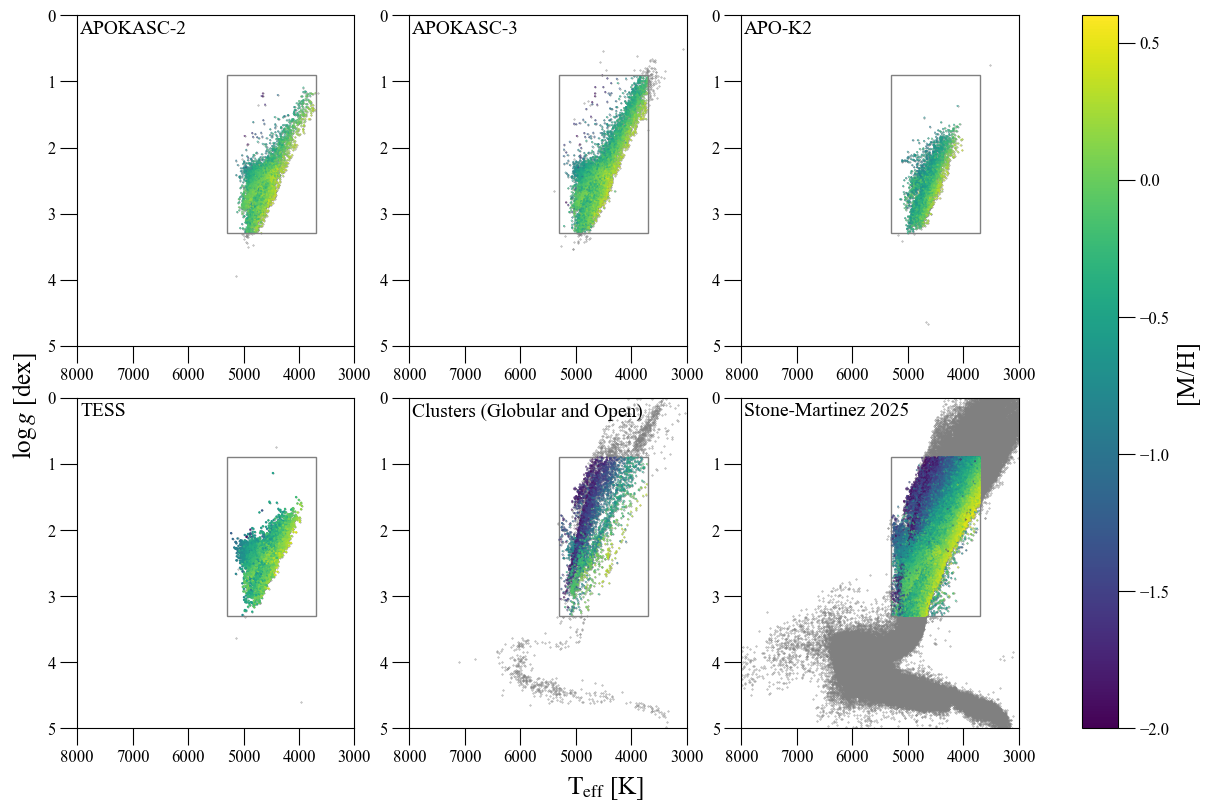}
    \caption{Kiel diagrams of our training, calibration, and comparison samples, color coded by metallicity. The grey box indicates the regime of interest, defined by the training data, for our analysis compared to each sample.}
    \label{fig:seismos}
\end{figure*}

In order to train our neural network to estimate ages, we need a set of stars of known ages. Here, in order to explore how sensitive our inferred ages are to the training sample, we use four different sets of stars where global seismic parameters such as the frequency of maximum power, \numax, which scales with the surface gravity, and the large frequency separation, \dnu, which scales with the mean density, have been measured. These values are used to estimate the stellar mass and surface gravity, and spectroscopy is used to estimate the metallicity. These components are then combined to estimate an age using a stellar model; previous work has indicated that this methodology is relatively insensitive ($\lesssim 10\%$) to the assumed physics and calibration of the underlying models \citep{Morales2025}. 

Our first sample is the APOGEE-Kepler 2 \citep[APOKASC-2,][]{Pinsonneault2018} sample, with {5,993} stars in our region of interest. This set combines results from six different seismic pipelines to estimate the seismic parameters, combines them with spectroscopic parameters from APOGEE DR 14 \citep{DR14}, infers evolutionary states as described in \citet{Elsworth2019} to apply theoretically-motivated corrections to \dnu\ \citep{Sharma2016}, and uses the open clusters in the Kepler field to correct \numax. These results have been used extensively in the literature as a training sample for machine-learning based age estimation \citep[e.g.][] {Ness2016, Mackereth2019}. 

Our second sample is the updated APOGEE-Kepler 3 sample \citep[APOKASC-3,][]{Pinsonneault2025} of {11,030} stars that meet our criteria. This work adds additional seismic pipelines to the analysis, remeasures the evolutionary states \citep{Vrard2025}, uses spectroscopic parameters from Data Release 17 \citep{DR17}, applies theoretically-motivated corrections to \dnu\ \citep{StelloSharma2022}, and calibrates \numax\ through a comparison to the radii from Gaia Data Release 3 \citep{GaiaDR3}. 

Our third sample is the APOGEE-K2 sample \citep[APO-K2,][]{SchonhutStasik2024} of 6292 stars within our region of interest. These have seismic results for \numax\ and \dnu\ from six pipelines as described in \citet{Zinn2022}, and spectroscopic parameters from APOGEE DR17 \citep{DR17}. The seismic parameters are corrected using theoretically motivated corrections based on spectroscopically inferred evolutionary states \citep{SchonhutStasik2024} as well as comparisons to the Gaia radii \citep{GaiaDR3}. Ages are computed from the seismic and spectroscopic parameters using models as described in \citet{Warfield2024}. 

Our fourth and final set of ages comes from the overlap between APOGEE and TESS \citep[TESS,][]{Theodoridis2025}, a set of {15,517} in our region of interest. These stars have \numax\ values inferred by applying a convolutional neural network to individual sectors of TESS data \citep{Hon2021}. Because they do not necessarily have measured \dnu\ values, radii were computed using Gaia DR2 \citep{GaiaDR2}, and metallicities and temperatures were taken from the Gaia XGBoost analysis \citep{Andrae2023}. Evolutionary states were estimated using the offset from stellar models, empirical mass loss assumptions were made following the results from \citet{Pinsonneault2025}, and ages were computed \citep{Theodoridis2025}. There is significant overlap between this sample and APOGEE \citep{TheodoridisTayar2023} and so we can use the {15,517} stars in this sample that are also in DR19 and meet our criteria as our final training set. We show the distribution of each of these training sets in Figure \ref{fig:seismos}.

\subsection{Validation Samples}\label{ssec:methodvalidation}
In order to determine the validity of the ages inferred from our network, we need a set of stars whose ages are already constrained by external information. For this we use stars in clusters. The absolute ages of these systems require significant work to constrain \citep[e.g.][]{Ying2023, Ying2024, Reyes2024, Ying2025, Reyes2025}, but it has long been agreed that such systems should be treated as coeval populations, and therefore the ages inferred independently for the two stars should agree to within their uncertainties.

In particular, we use the {Open Cluster Chemical Abundances and Mapping} \citep[OCCAM,][]{Frinchaboy2013} sample of cluster members determined for DR19 \citep{Otto2026}. These stars are identified as potential cluster members using parallaxes, proper motions, and positions from Gaia \citep{Cantat-Gaudin2020, Hunt2023} and then their membership is spectroscopically confirmed using APOGEE abundances and radial velocity values. This provides us a list of {637 giants in 94 open clusters}
that we can use to evaluate our age results. In order to broaden the age and metallicity ranges of our calibrators, we also include {4,217 stars in 20 globular clusters} identified using {similar techniques} in APOGEE by \citet{Schiavon2024}. While we considered adding co-moving likely wide binary systems identified in Gaia \citep{ElBadry2021} to our sample, we found only one result where both stars had spectroscopic results from APOGEE and both had red giant parameters consistent with our sample cuts; we therefore decided that this sample was too small at this time to provide useful verification. The properties of our cluster validation samples are also shown in Figure \ref{fig:seismos}.

\subsection{Sample Selection} \label{ssec:methodselect}
Machine learning methods generally perform best on data similar to what is in their training sets. Therefore, in order to fairly evaluate the resulting ages, we restrict our sample to stars with similar properties to the red giant oscillators described in Section \ref{ssec:Seismic}. In particular, we restrict our sample to the {351,995} stars with {inferred ages larger than $0$ $\mathrm{Gyr}$}, temperatures between $3700$ and $5300$ $\mathrm{K}$, log(g) values between $0.9$ and $3.3$ dex, metallicities between $-2.0$ and $0.6$ dex. We also remove stars with `bad' flags {or whose $\mathrm{[C/H]}$, $\mathrm{[M/H]}$, $\mathrm{[N/H]}$ measurements have undefined or infinite values}, since previous work \citep[e.g.][]{Meszaros2025} suggests that these stars will not have reliable parameters.

\begin{figure}[hb]
    \centering  
    \includegraphics[width=\linewidth]{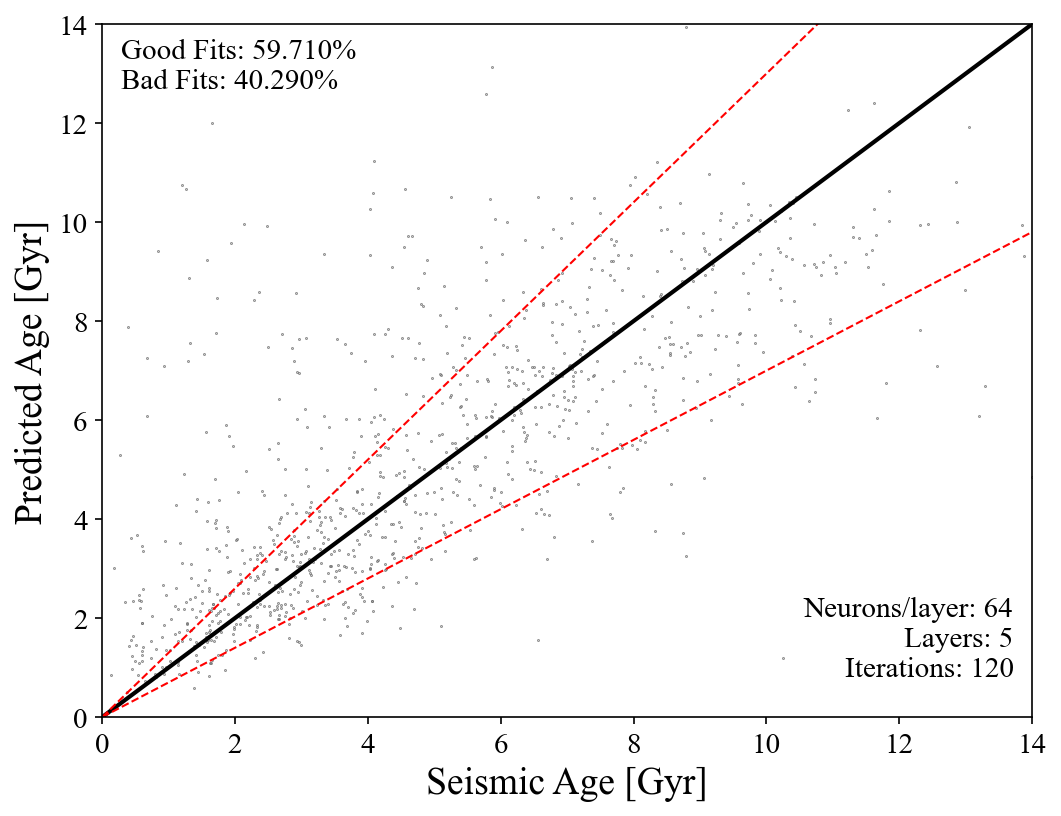}
    \includegraphics[width=\linewidth,clip=true, trim=0in 0.3in 0in 0.3in]{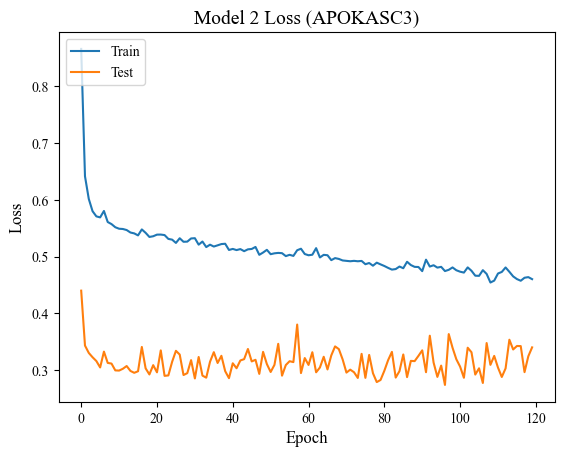}

    \caption{\textbf{Top:} Recovery plot for our preferred base case neural network, Model 2 trained on APOKASC-3 data. Points indicate the predicted versus seismic ages individual stars in the test set, with the black line representing agreement and the dotted red lines marking offsets of 30 percent. \textbf{Bottom:} Loss plot for Model 2 trained on APOKASC-3 data for both the training (blue) and testing (orange) sets. While loss continues to decrease for the training set, it has stabilized for the test set indicating that a sufficient number of iterations has been reached.}  
    \label{fig:lossplot}
\end{figure}

\subsection{Neural Network} \label{ssec:methodml}
Following the methodology described in \citet{Tayar2023RNAAS}\footnote{https://zenodo.org/records/10392858}, we use \texttt{TensorFlow keras} \citep{tensorflow2015-whitepaper} to construct a simple, fully-connected neural network with linear output, mean squared error loss, ReLU activation \citep{ReLU}, and an Adam optimizer \citep{Adam}. We explore in Section \ref{ssec:varyhyper} the impact of the number of nodes per layer, the number of layers and the number of iterations in training (see Table \ref{tab:model_params}). Our network takes as inputs the temperature $\mathrm{T_{eff}}$, surface gravity log(g), metallicity [M/H], $\alpha$-element abundance [$\alpha$/M], carbon abundance [C/H], and nitrogen abundance [N/H], trains on one of the age samples described in Section \ref{ssec:Seismic} and returns an age. This is substantially simpler than many machine learning methodologies used to estimate ages in the literature \citep[e.g.][]{Mackereth2019, Leung2023, StoneMartinez2025}, and we describe in Section \ref{ssec:mltypevary} the impact of this simplicity. However, the relationships between temperature and age on the giant branch \citep[e.g.][]{SandersDas2018, Morales2025} as well as the relationship between the [C/N] and age \citep{Martig2016, Roberts2024, Spoo2025} generally seem to have relatively simple shapes, and so it is conceivable that such a simple structure may be sufficient to make reliable age predictions. We show in Figure \ref{fig:lossplot} an example where the model reasonably reproduces the ages in the training set, and seems converged. We also make available the code used for this paper on Github \footnote{https://github.com/jtayar/2025\_AST4300\_Project3} and Zenodo \footnote{https://zenodo.org/records/20317619}.

\begin{table}[h]
    \centering
    \footnotesize
    \begin{tabular}{|c|c|c|c|c|}\hline
         \textbf{Model}&  \textbf{Neurons/Layer}&   \textbf{Layers}& \textbf{Iterations}&\textbf{Ages $<$ 30\%}\\\hline
         \textbf{1}&  32&  10& 150 &59.034\%\\\hline
         \textbf{2}&  \textbf{64}&  \textbf{5}& \textbf{120} &\textbf{59.710\%}\\\hline
         \textbf{3}&  20&  5& 200 &57.005\%\\\hline
         \textbf{4}&  20&  5& 5000 &55.652\%\\\hline
         \textbf{5}&  100&  5& 200 &60.386\%\\\hline
         \textbf{6}&  20&  10& 200 &57.000\%\\\hline
         \textbf{7}&  100&  10& 1000 &57.198\%\\ \hline
    \end{tabular}
    \caption{We train a variety of neural networks with a range of layers, neurons per layer, and iterations. We report in the final column the percent of APOKASC-3 test set ages predicted to within 30\% accuracy. In general we find that the network hyperparameters do not significantly affect our results, but we chose Model 2 as our preferred base case. }
    \label{tab:model_params}
\end{table}

\section{Analysis} \label{sec:analysis}
While predicting ages using machine learning is relatively straightforward, we are interested in examining more specific questions about the ages produced by this technique. In particular, we want to know how sensitive the ages are to the hyperparameters of the neural network (Section \ref{ssec:varyhyper}), how sensitive the predicted ages are to the training set chosen (Section \ref{ssec:trainingvary}), and how well our ages compare to those predicted using more sophisticated machine learning methodologies and architectures (Section \ref{ssec:mltypevary}). We also compare our predicted ages to other seismic results as well as cluster analyses to infer which methodologies are producing the most accurate ages (Section \ref{ssec:agevalidation}) and in what parts of parameter space are the results most sensitive to these choices (Section \ref{ssec:agedifferences}). In general, we find that our results are not particularly sensitive to the chosen model neurons, layers, or iterations, that they are generally consistent with the predictions of more sophisticated models from the literature, and that the predictions are most different in regimes with limited training data, including high-luminosity stars, low-metallicity stars, and low-temperature stars. Given our comparisons in this section, our preferred model is Model 2 (See Table \ref{tab:model_params}), trained on the APOKASC-3 data \citep{Pinsonneault2025}, and so we use that for our base case for comparison and presentation, although we have cross-compared a variety of different options in the course of our exploration.

\subsection{Network Variation} \label{ssec:varyhyper}

In our exploration, we varied the number of layers, the number of nodes per layer, and the number of iterations. Consistent with previous work \citep[e.g.][]{Tayar2023RNAAS}, we find that these variations do not significantly alter our fraction of recovered ages as long as the model has sufficient iterations to converge and is not strongly overfit. For our later analysis (see Section \ref{ssec:agemaps}), we have used the results from Model 2, which has a high recovery rate and a simpler network, with loss plots and validation against the testing data that indicated the model was well-fit but not overfit (Figure \ref{fig:lossplot}). 
%
%

\begin{figure}[h]
    \centering
    \includegraphics[width=0.99\linewidth]{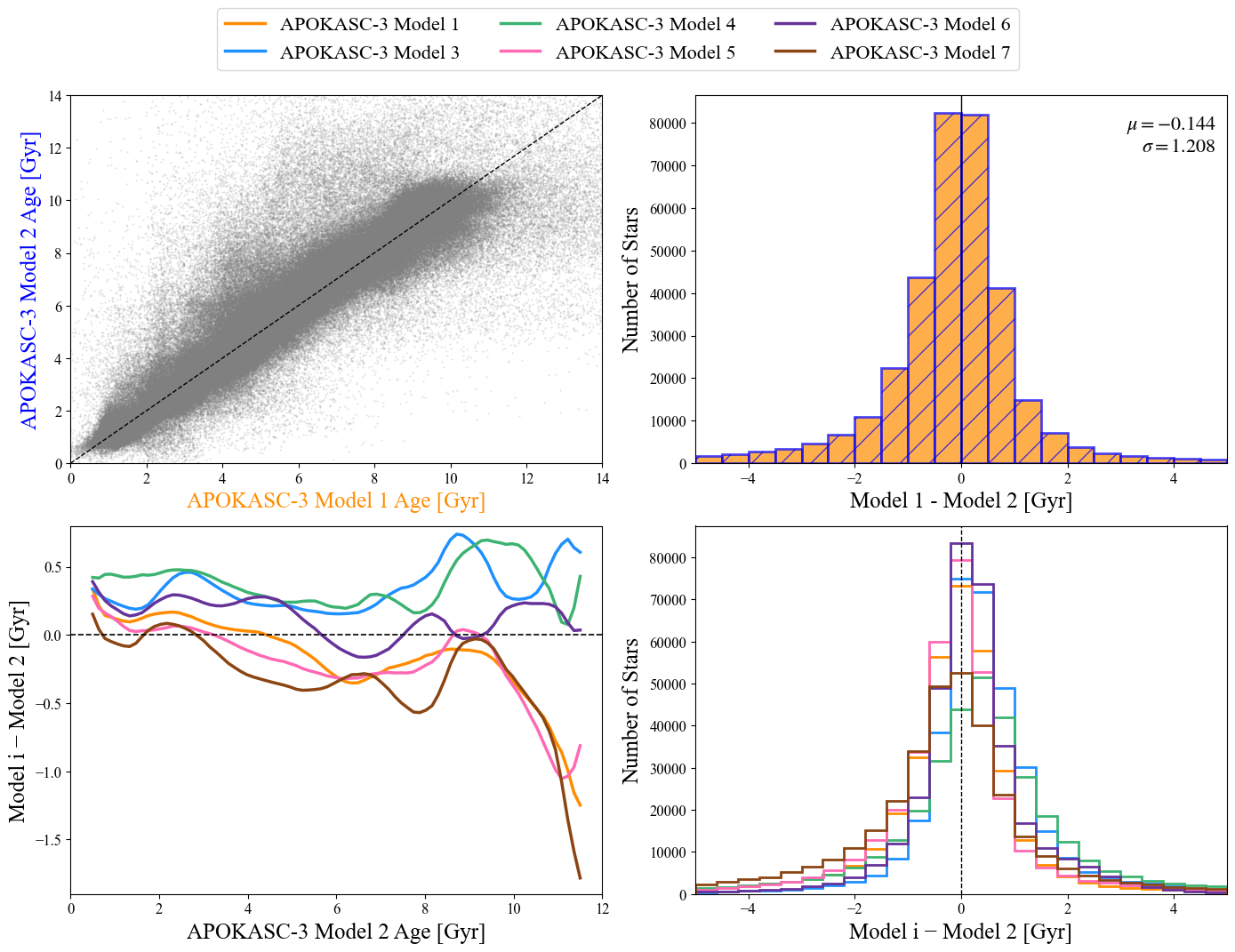}
    \caption{Comparison of age predictions from models with different layers, neurons per layer, and iterations relative to our baseline Model 2. \textbf{Top:} Direct comparison of Model 1 and Model 2 is shown in the top left plot, with the corresponding residual distribution in the top right. \textbf{Bottom:} Running median residuals (left) and residual histograms (right) for Models 1 (orange), 3 (light blue), 4 (green), 5 (pink), 6 (purple), and 7 (brown) relative to Model 2 (blue). All models show consistent age predictions (offsets $<0.3$ Gyr, scatter $<1.5$ Gyr) except at old ($>9$ Gyr) ages.}
    \label{fig:networkvariations}
\end{figure}

We show the direct comparison for two sets of models in Figure \ref{fig:networkvariations} (top) and a more generalized comparison of a set with varied hyperparameters to our preferred, well-trained model (Figure \ref{fig:networkvariations}, bottom). In general we find only slight differences between models all trained on the APOKASC-3 data with different numbers of layers, iterations, and nodes, with an average offset of less than 0.3 Gyrs and scatter $<$1.5 Gyrs, and we found similar internal consistency when training on other datasets as well. We do note that larger differences between models start to appear at the oldest ages, where there is less training data \citep{Pinsonneault2025}, and the [C/N] ratio becomes less sensitive to age \citep{Roberts2024}. The challenges of constraining the oldest ages using machine learning on spectroscopic samples like Milky Way Mapper have been previously documented \citep[e.g.][] {Mackereth2019, FAnders2023} and we discuss it further in Section \ref{ssec:agedifferences}.

\subsection{Training Set Variations} \label{ssec:trainingvary}

\begin{figure}[h]
    \centering  
    \includegraphics[width=\linewidth,clip=true, trim=0in 0in 0in 0in]{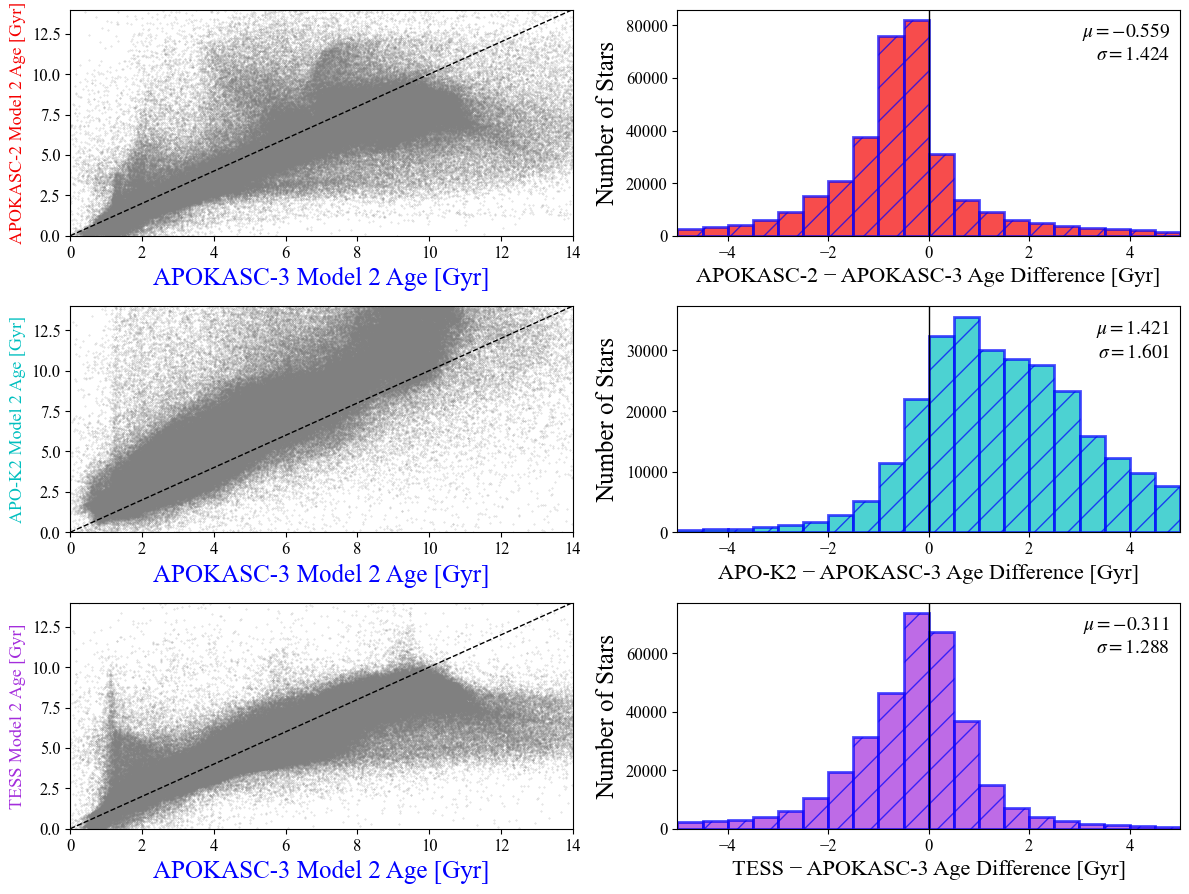}
    \caption{Demonstration of the impact of changing the training dataset to APOKASC-2 (top, red), APO-K2 (middle, teal), or TESS (bottom, purple) compared to the base case ages inferred by training on APOKASC-3 (blue). When computing the mean and standard deviation for each sample, we have removed significant outliers ($|\Delta {\rm Age} |> 5$ Gyr).} \label{fig:varytraining}
\end{figure}

Our next question concerns the impact of the training set on the inferred ages. We show in Figure \ref{fig:varytraining} the comparison of the ages inferred using the other three training sets (APOKASC-2, APO-K2, and TESS) compared to the ages inferred when training on APOKASC-3, all using the hyperparameters for Model 2. For the APOKASC-2 and TESS sets, we see an offset of $\sim 0.5$ Gyr, and a scatter of $<1.5$ Gyr. This suggests that the chosen training set generally adds a small additional offset but an insignificant additional scatter when compared to the random fluctuations between the machine learning networks. We do, however, see an additional offset (1.5 Gyr) and a somewhat larger scatter when training on the APO-K2 data, suggesting a systematic shift between the APO-K2 data and the other training datasets.

\subsubsection{APO-K2} \label{sssec:apok2}
\begin{figure}[H]
    \centering
    \includegraphics[width=0.87\linewidth]{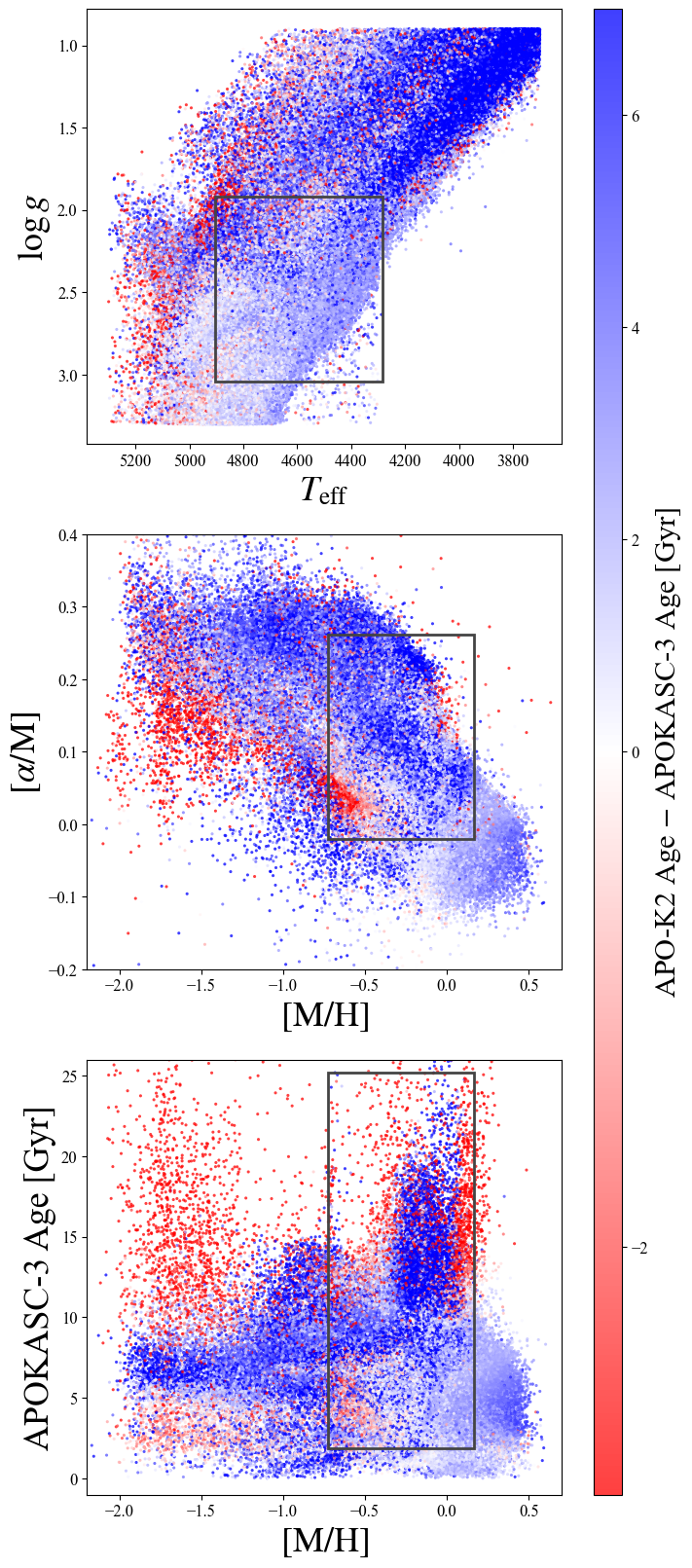}
    \caption{Residuals between the predictions of the network trained on APOKASC-3 and the network trained on APO-K2 in a Kiel diagram (top), chemical space (middle), and age-metallicity space (bottom). We mark with a gray box the location in each plot of the majority of the APO-K2 training data, and find significant offsets in the results (identified as bright red or bright blue points) far from this region.}
    \label{fig:apok2_param_scatter}
\end{figure}

We have done some additional exploration with the APO-K2 data to try to understand this offset. In particular, we have 
1.) tried using only the RGB stars, since clump stars were treated differently in the \citet{Warfield2024} analysis, and 2.) double checked our training data and varied the network parameters. 
Our best understanding is that in addition to a small average shift between ages inferred for similar stars between APO-K2 and APOKASC-3 \citep{Warfield2024}, the main problem is that the APO-K2 training dataset is small, and covers a relatively restricted parameter space, with the majority of the training set falling within the grey boxes in Figure \ref{fig:apok2_param_scatter}. Therefore, when used to train a network that is applied to a wide range of red giants, it does not accurately extrapolate to lower mass stars, high-$\alpha$ stars, metal-rich stars, and luminous giants, and this significantly biases the average ages of the full Milky Way Mapper DR19 sample. We therefore recommend that future users be cautious with analyses over wide parameter spaces that are primarily trained on the APO-K2 dataset.   

\subsection{Comparison To Other Machine Learning Methodologies} \label{ssec:mltypevary}

\begin{figure}[tb]
    \centering  \includegraphics[width=0.47\textwidth,clip=true, trim=0in 0in 0in 0in]{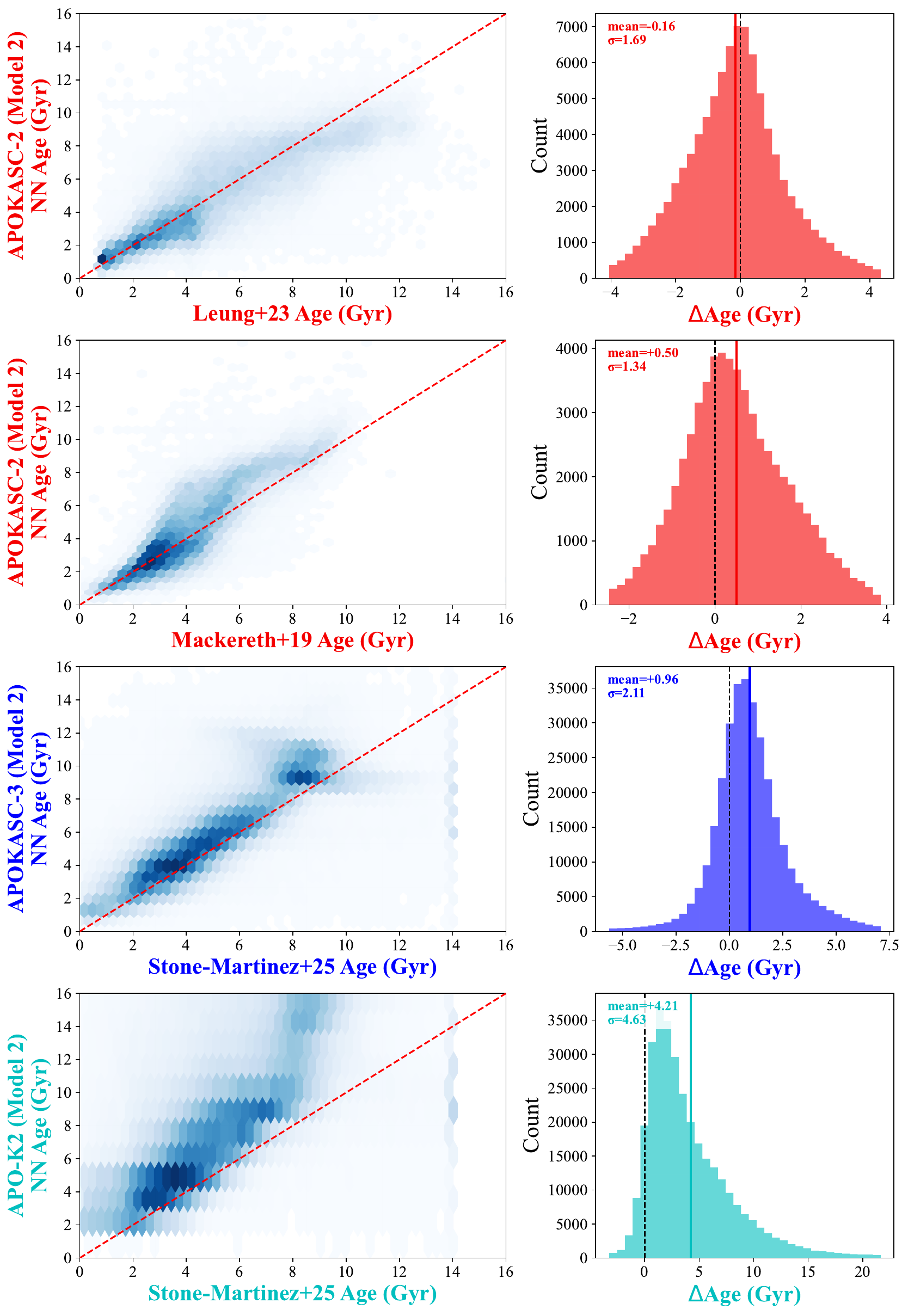}
    \caption{\textbf{Left:} Comparison of our predicted ages to literature ages inferred using more sophisticated machine learning techniques (top, variational encoder-decoder, \citealt{Leung2023}; second from top, Bayesian Convolutional Neural Networks, \citealt{Mackereth2019}; bottom two rows, normalizing flows, \citealt{StoneMartinez2025}) but similar training sets (see text). \textbf{Right:} Histograms, colored by the training set (APOKASC-2 in red, APOKASC-3 in blue, APO-K2 in teal), of the differences between our predictions and the literature predictions generally show consistent results with small offsets but some scatter.} 
    \label{fig:varyarchitecture}
\end{figure}
 
While our chosen machine learning methodology is an extremely simple, fully-connected neural network, more sophisticated architectures have been applied to this problem of estimating ages for large numbers of red giants using spectroscopic parameters inferred from the APOGEE spectrograph, and we want to determine whether these more sophisticated approaches make significantly different predictions. In Figure \ref{fig:varyarchitecture} we show comparisons between our baseline Model 2, and the predictions from \citet{Leung2023}, which uses {a variational encoder-decoder network}, \citet{Mackereth2019}, which uses {a Bayesian Convolutional Neural Network} and \citet{StoneMartinez2025}, which uses {normalizing flows}. 

To reduce the offsets between methods from the training set differences, we try to match the training data choices used in those papers for our comparison. In particular, \citet{Leung2023} trained on the {\citet{Miglio2021a} reanalysis of the APOKASC-2 data}. For consistency with our other comparisons, we use Model 2 trained on APOKASC-2 for the comparison to the \citet{Leung2023} results, but we have verified that changing to the \citet{Miglio2021a} dataset does not have a significant impact on our comparison. For the comparison to \citet{Mackereth2019}, we also compare to Model 2, trained on the APOKASC-2 dataset. Finally, for the \citet{StoneMartinez2025} analysis, the authors trained on a combination of the APOKASC-3 and APO-K2 datasets, and so we show comparisons to each of our Model 2 network predictions trained on these datasets. We see the offset described in Section \ref{sssec:apok2}, so we focus on the comparison to the model trained on the larger APOKASC-3 sample. 

In each of these cases, we see consistent results between our predictions and the more sophisticated literature models, with average offsets of {$\sim$0.5 Gyr and a scatter of $\sim$2 Gyr}. This is {consistent with} the offsets we see between networks using different hyperparameters but the same training set, described in Section \ref{ssec:varyhyper}, but has a slightly larger scatter, suggesting that the choice of machine learning methodology {has a small impact} for this particular problem, inferring ages from spectroscopic parameters for red giants. Given that the simplicity of our neural network does not substantially bias the predicted ages within this parameter space, it seems likely that future improvements in age inference will depend on improving and expanding the training sets rather than increasing the complexity of the technique or architecture.

\subsection{Age Validation} \label{ssec:agevalidation}

\begin{figure}[tb]
    \centering  \includegraphics[width=0.47\textwidth,clip=true, trim=0in 0in 0in 0in]{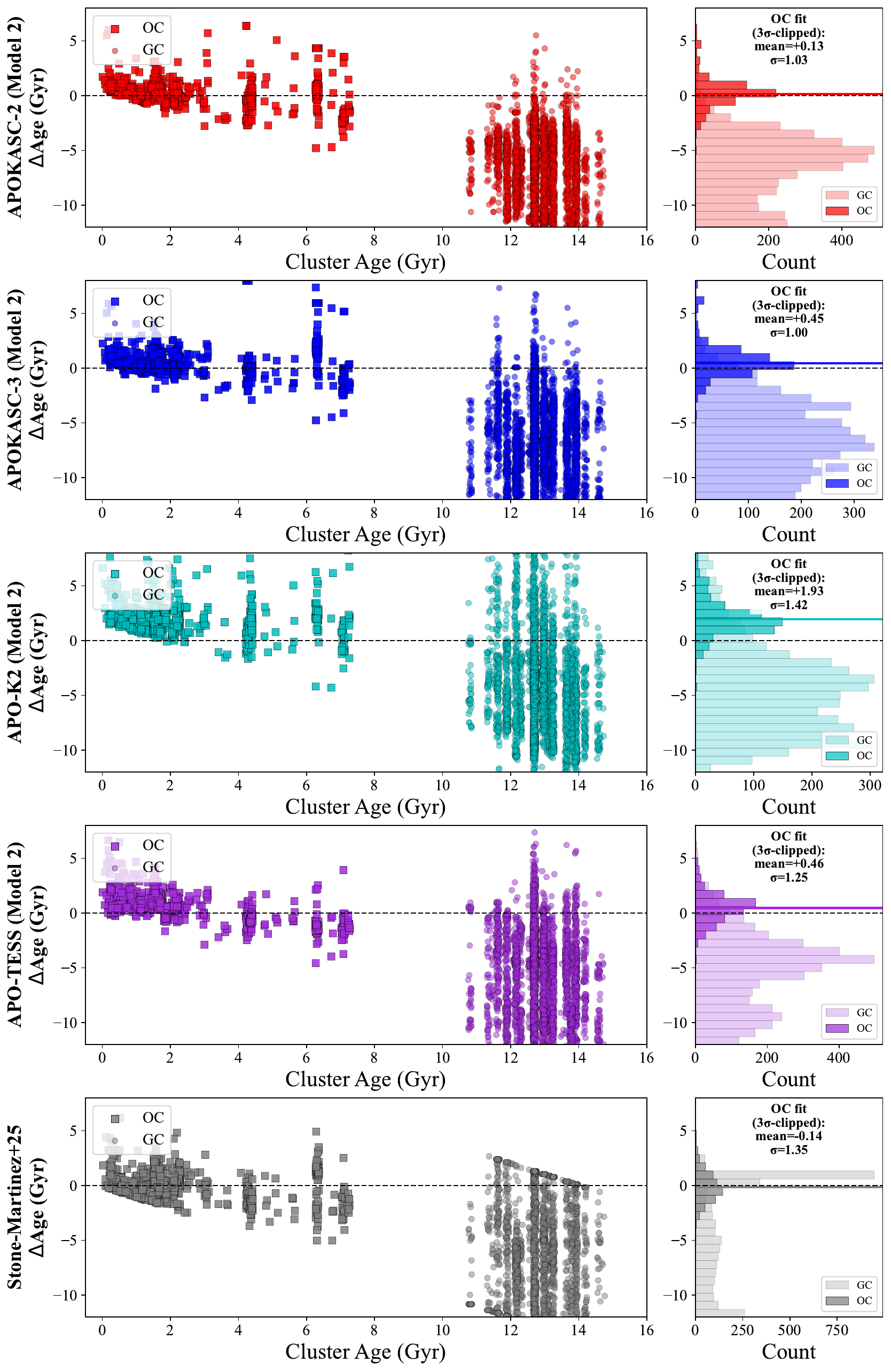}
    \caption{Differences between the ages predicted from networks trained on each sample (from top: APOKASC-2, red; APOKASC-3, blue; APO-K2, teal; TESS, purple) and age of the clusters for each member of an open (squares and opaque histograms) or globular (circles and translucent histograms) cluster. We also show the comparison between the published \citet{StoneMartinez2025} ages and the cluster ages in the bottom row. While ages for stars in open clusters are generally predicted well, with limited scatter, predictions for globular cluster stars are significantly offset from their published ages.} 
    \label{fig:clustervalidation}
\end{figure}

\begin{figure}[tb]
    \centering  \includegraphics[width=0.5\textwidth]{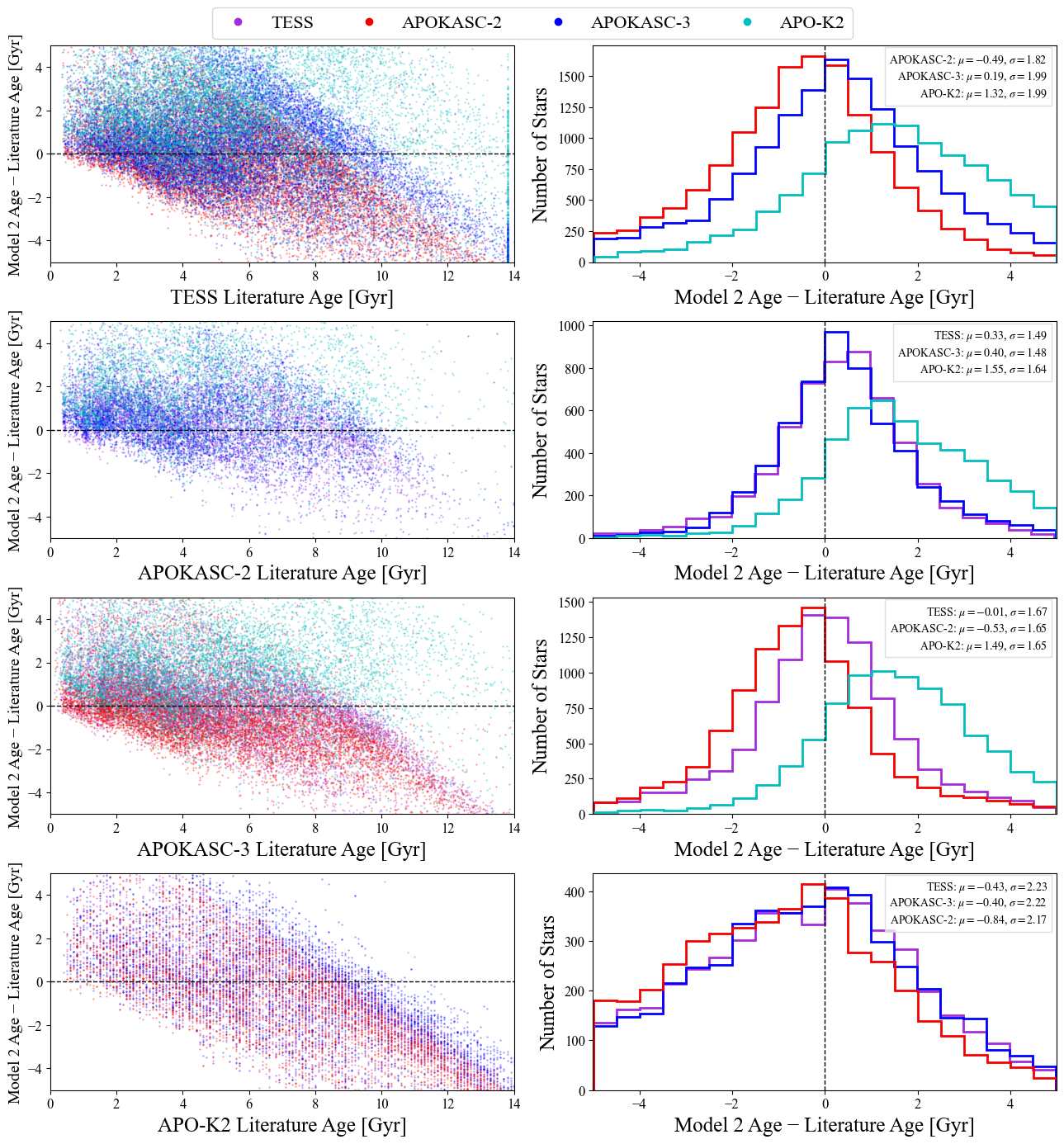}
    \caption{Offsets between the ages predicted by our networks and the seismic ages for values that were not included in the network's training set (colors: APOKASC-2, red; APOKASC-3, blue; APO-K2, teal; TESS, purple). Offsets between the measured and predicted values are generally small for stars younger than 9 Gyr.} 
    \label{fig:seismicvalidation}
\end{figure}

While the previous sections discussed the offsets and scatter between age prediction methods, it is also useful to consider the accuracy of the predictions. For stars in clusters, we expect that all of the stars in the same cluster should be the same age, and so the inferred age scatter within a cluster is a measure of the systematic shifts in the inferred ages as a function of position on the Hertzsprung-Russell Diagram. Additionally, we can compare our results to the literature ages for these clusters to get a sense of which training set produces the most reliable ages. We show these comparisons in Figure \ref{fig:clustervalidation}. In general, we find that the ages predicted for open cluster stars using the TESS, APOKASC-2, and APOKASC-3 training samples are quite good, with average offsets of $\lesssim$ 0.5 Gyr and a scatter of $\sim$ 1 Gyr. The scatter is consistent with or lower than our model-to-model and training set-to-training set expected offsets, although the average offset is slightly larger, hinting at potential systematics between seismic and cluster age scales \citep{TayarJoyce2025}. For the ages trained on APO-K2, we see the expected offset of $\sim$ 2 Gyr (see Section \ref{sssec:apok2}) in the predicted ages, although the scatter is relatively small. Our comparison to cluster data indicates that training on APOKASC-2, which was calibrated using clusters, gives the ages most consistent with the cluster scale, although training on APOKASC-3 or TESS produce results that are almost as accurate and precise. 

We also note that none of our networks produce reliable age estimates for globular cluster stars. Part of this may be related to the challenges of inferring reliable ages for old stars \citep[e.g.][]{Mackereth2019, StoneMartinez2025} from methods like this. However, globular clusters also have second generation abundance patterns \citep{Cohen1978, BastianLardo2018}, extra mixing \citep{Gratton2000, Kirby2016, Shetrone2019, TayarJoyce2022}, and other complexities that can complicate age inference \citep{Spoo2025}. While there have been some efforts to directly infer asteroseismic ages from globular cluster stars \citep[e.g.][]{Tailo2022, Howell2022, Howell2024, Howell2025}, additional work is likely needed \citep{HAYDN} to provide additional age calibrators in such systems and at the oldest ages in general.

We also have seismic ages for tens of thousands of stars in this sample. While we must exclude from the comparison the ages that were used to train each network, we can explore how well a network trained on one set of data (e.g. APOKASC-3) predicts results for ages it has not seen (e.g. from TESS). We show in Figure \ref{fig:seismicvalidation} the offset between the predicted age and the seismic age for each network, using the Model 2 hyperparameters. We find the expected offset for the APO-K2 sample, as well as a slight systematic shift between APOKASC-2 and APOKASC-3, which may result from the change in calibration methodologies between these samples. In general, the networks trained on APOKASC-3 and TESS both do a very good job of predicting the other seismic age measurements on which they were not trained, with offsets of $<0.5$ Gyr and a scatter of $<2$ Gyr, only slightly larger than the network-to-network and training set-to-training set variations. Given the cluster results, seismic comparisons, as well as previous work, we choose APOKASC-3 as our baseline for comparison, but we find the results inferred from training on the TESS or APOKASC-2 data almost equivalently reliable.

\subsection{Understanding Age Differences} \label{ssec:agedifferences}
\begin{figure}[tb]
    \centering  \includegraphics[width=0.47\textwidth,clip=true, trim=0in 0in 0in 0in]{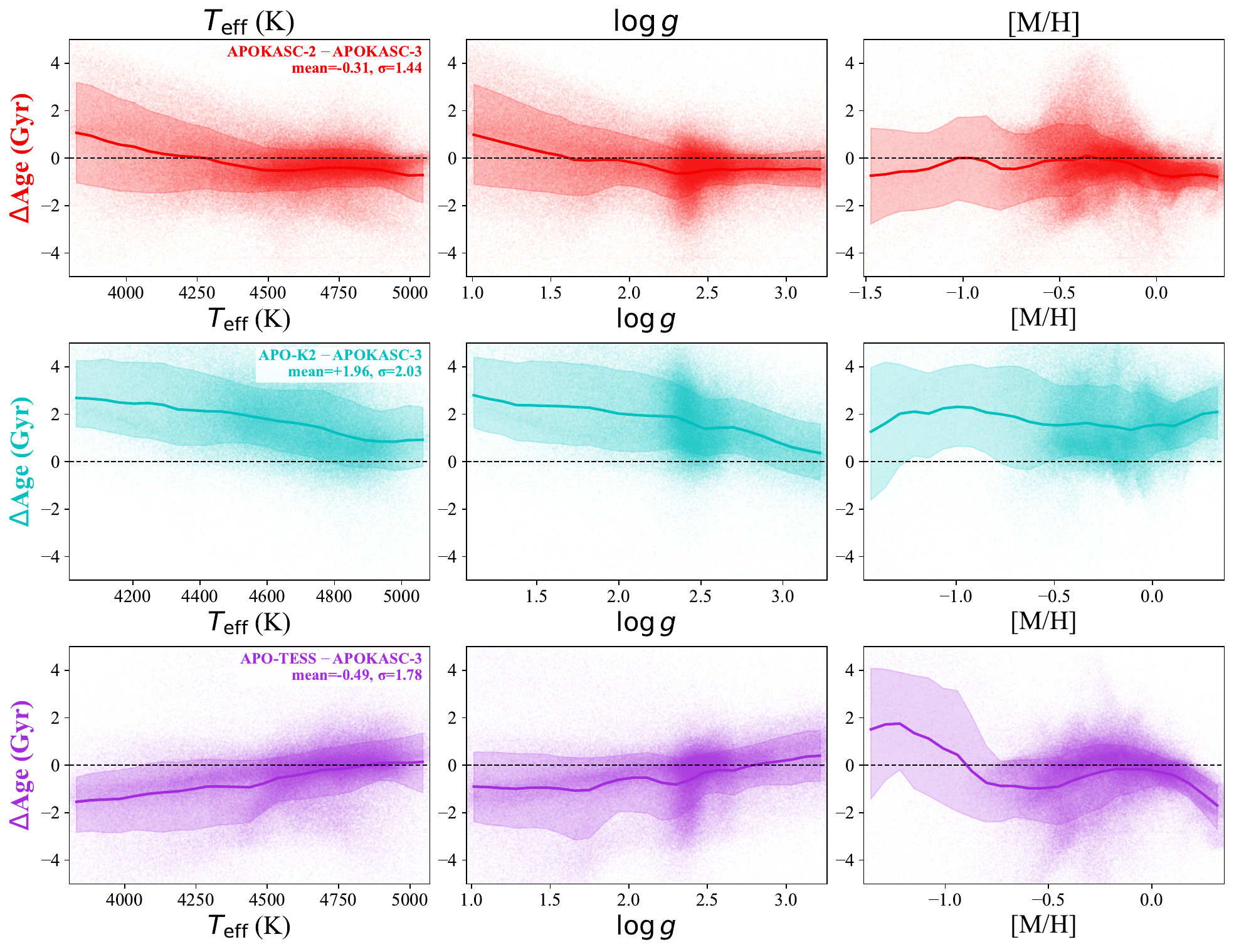}
    \caption{Offsets between the age predictions of models trained on APOKASC-2 (top, red), APO-K2 (middle, teal), and TESS (bottom, purple) compared to our baseline model trained on APOKASC-3 as a function of temperature (left), surface gravity (middle), and metallicity (right). For each plot, we have marked a running median (solid line) and uncertainty band (16-84th percentile) to emphasize regimes where predictions differ depending on the chosen model, such as low temperature, low gravity, and low metallicity.} 
    \label{fig:subsetstraining}
\end{figure}

\begin{figure}[tb]
        \centering
        \includegraphics[width=0.47\textwidth,clip=true, trim=0in 0in 0in 0in]{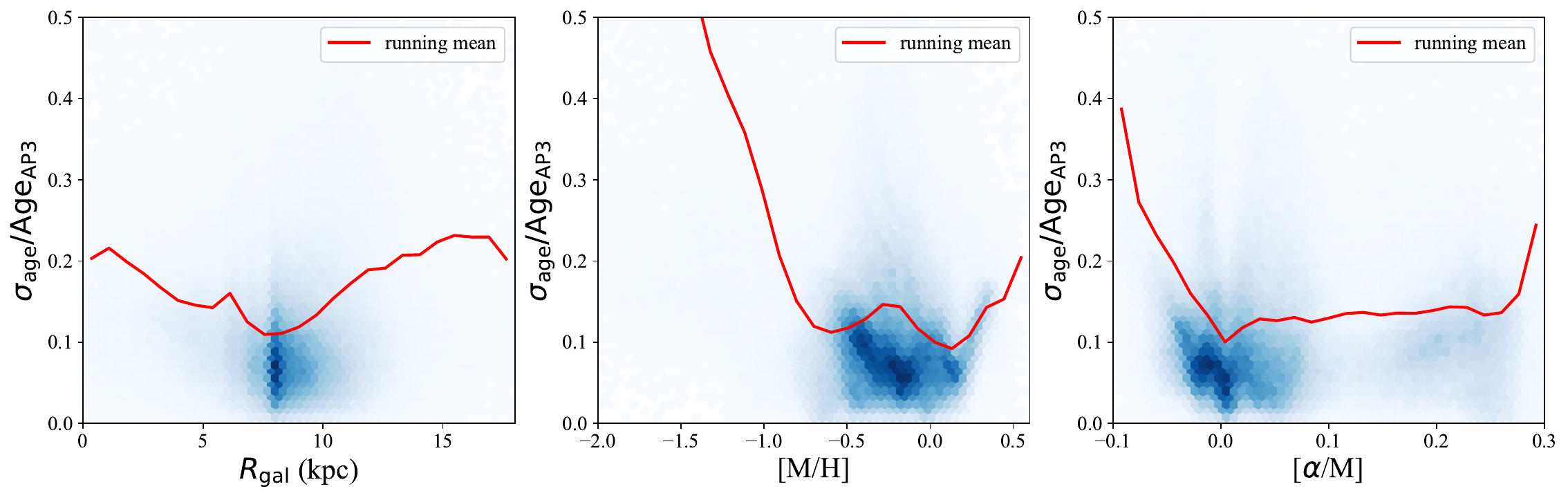}
        \caption{Average fractional variation in age between models trained on the APOKASC-2, APOKASC-3, and TESS samples as a function of galactocentric radius (left), metallicity (center), and $\alpha$-element enhancement (right). In each panel, the color indicates the density of points, with darker regions having more points, and we have shown the running median as a red line.}
        \label{fig:sensitivesamplesparamsfrac}
\end{figure}

\begin{figure}[tb]
        \centering
        \includegraphics[width=0.47\textwidth,clip=true, trim=0in 0in 0in 0in]{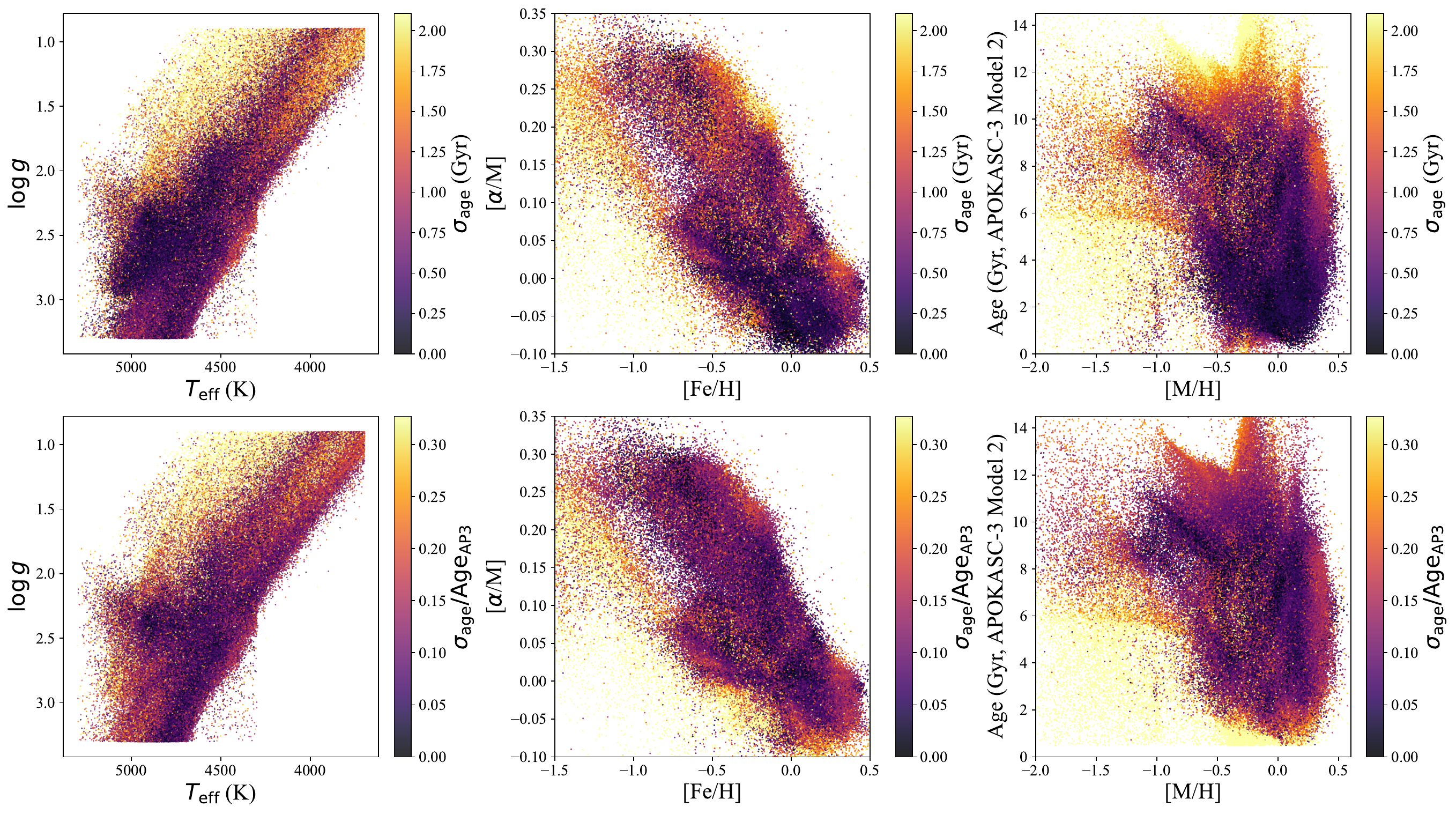}
        \caption{Absolute (top) and fraction (bottom) scatter in ages predicted by models using different training sets (APOKASC-2, APOKASC-3, or TESS) for stars in our sample as a function of Kiel diagram position (left), composition (middle), or age and metallicity (right). Dark colors indicate ages that are consistently inferred while bright colors indicate significant systematic uncertainties in age.}
        \label{fig:sensitivesamplesplanes}
\end{figure}

\begin{figure}[tb]
        \centering
        \includegraphics[width=0.47\textwidth,clip=true, trim=0in 0in 0in 0in]{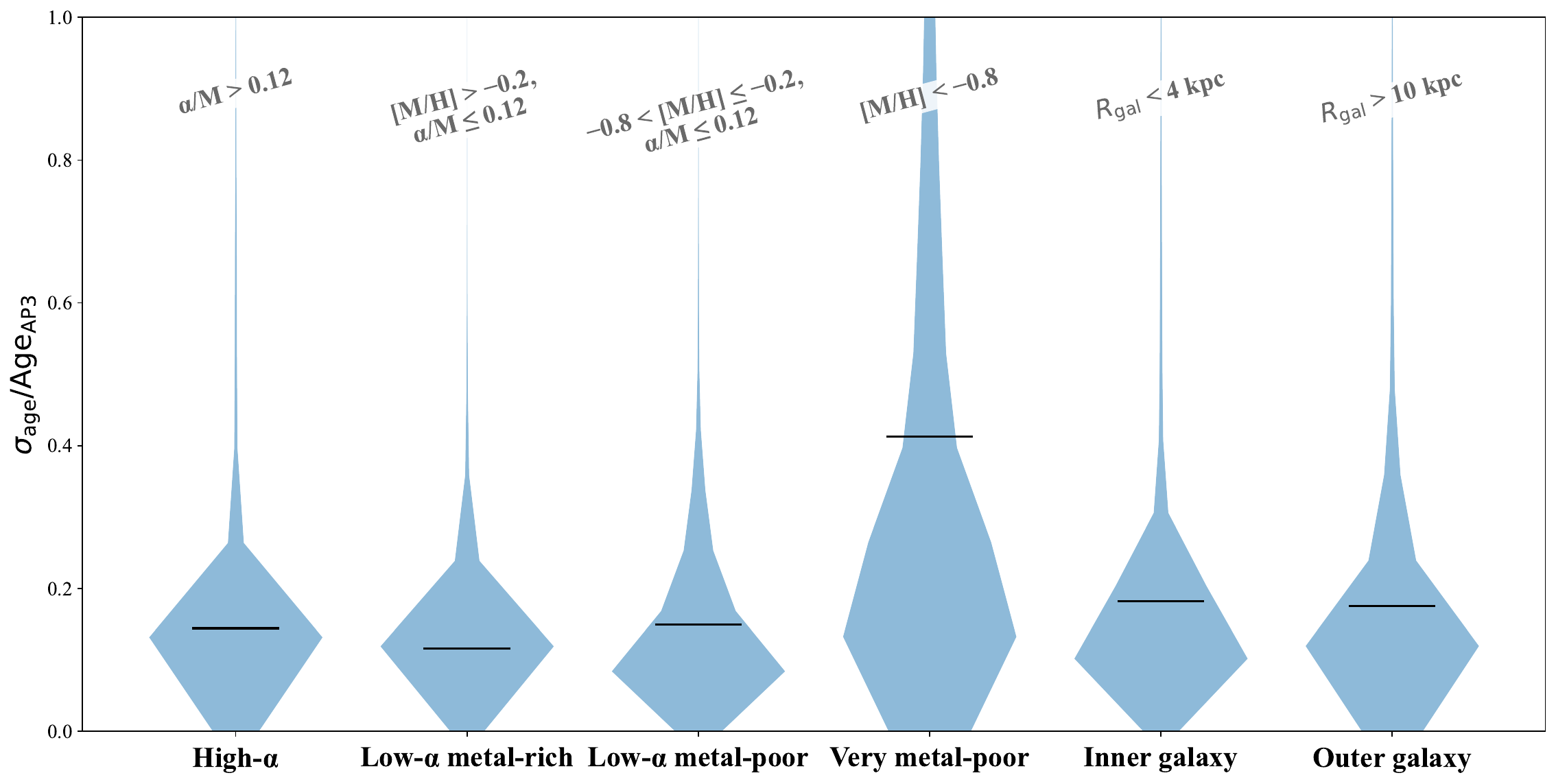}
        \caption{Violin plots showing the distribution of fractional uncertainties in age predicted by networks using different training sets (APOKASC-2, APOKASC-3, TESS) for subsets of our data. For each subset the average \jt{is this a mean or a median} value is marked with a black line.}
        \label{fig:sensitivesamplesviolinfrac}
\end{figure}

While the overall quality of the predicted ages is important, there are also a variety of interesting scientific explorations that must rely on ages for particular subsets of the data. In Figure \ref{fig:subsetstraining}, we show how the inferred ages vary when using different training sets as a function of temperature, gravity, and metallicity. While the majority of the stars in the sample are consistent between methods, stars at low temperatures, low gravities, and low or high metallicities are significantly more likely to have ages that are method-sensitive, with differences of several Gigayears and fractional offsets of tens of percent (Figure \ref{fig:sensitivesamplesparamsfrac}). This means that when looking at populations of stars (Figure \ref{fig:sensitivesamplesplanes}) the offsets between ages inferred using different methods will depend sensitively on the part of parameter space. We further break down these offsets as a function of population and parameters in Figure \ref{fig:sensitivesamplesviolinfrac}, which suggests that ongoing efforts to understand the evolution of the halo \citep{Chiappini2001, HortaSciavon2024}, the accretion of dwarf galaxies \citep{Helmi2018, Feuillet2020} and the growth of the bulge \citep{Griffith2021, Joyce2023, Miller2025} as well as the full {Galactic Genesis project \citep{Kollmeier2026}} may be significantly affected by limited training data and method-to-method systematics.  Recent efforts to detect and characterize seismic signals in luminous red giants from the ground \citep{Auge2020, Hey2023, Hey2025, Schochet2026} and predictions for seismic yields from the upcoming Roman \citep{Roman} Galactic Bulge Time Domain Survey \citep{Weiss2025} provide optimistic indications for a path forward in continuing to explore these exciting questions of galactic archaeology.



\section{Discussion} \label{sec:discussion}

In Section \ref{sec:analysis}, we have shown that we can estimate ages for a large sample of 351,995 stars from Milky Way Mapper Data Release 19, and that our ages are competitive with previously published analyses. Given these results, we make a few diagnostic plots of the inferred Milky Way evolution from this sample, with the goal of demonstrating the range of the sample, and encouraging future, more detailed, analyses that appropriately account for and correct for the selection effects inherent in this sample, and treat binaries \citep{Troup2016, PriceWhelan2020, Daher2022}, evolutionary state separation including the Asymptotic Giant Branch  \citep[AGB,][]{Vrard2025}, dwarf galaxy stars \citep{Naidu_2020, Shetrone2026} and other unique subsamples more carefully. 

\begin{figure*}[tbh]
    \centering  \includegraphics[width=0.99\textwidth,clip=true, trim=0in 0in 0in 0in]{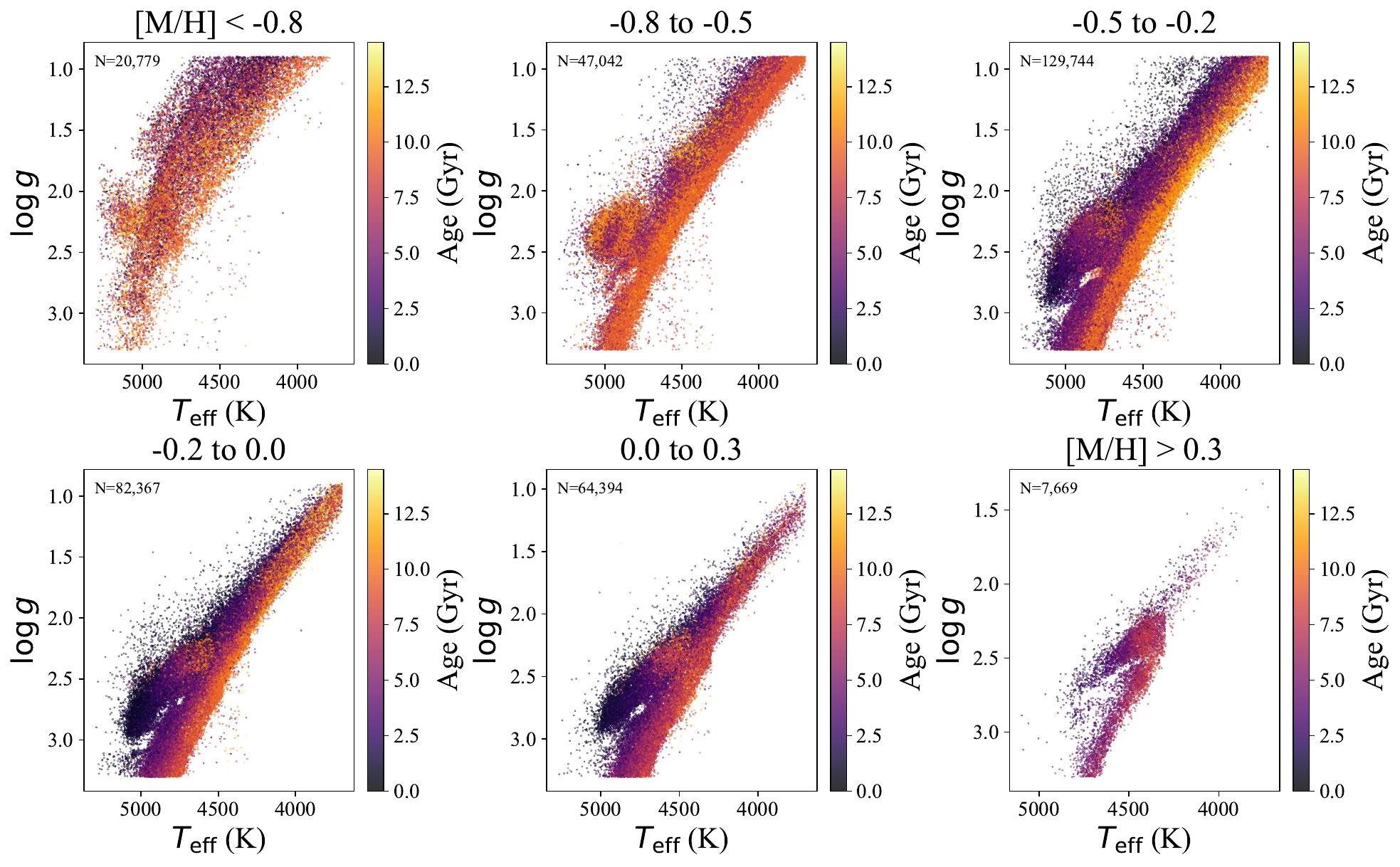}
    \caption{Kiel diagrams with each star color-coded by age for stars in each metallicity bin from the most metal poor (top left) to the most metal rich (lower right). In each case, old stars are marked as bright points and young stars are marked as dark points. In each panel, the expected trends are seen in age for both the shell-hydrogen-burning first-ascent red giant branch stars and the core-helium-burning clump stars. } 
    \label{fig:agemapkiel}
\end{figure*}

\subsection{Age Mapping} \label{ssec:agemaps}

\begin{figure*}[tbh]
    \centering  \includegraphics[width=0.99\textwidth,clip=true, trim=0in 0in 0in 0in]{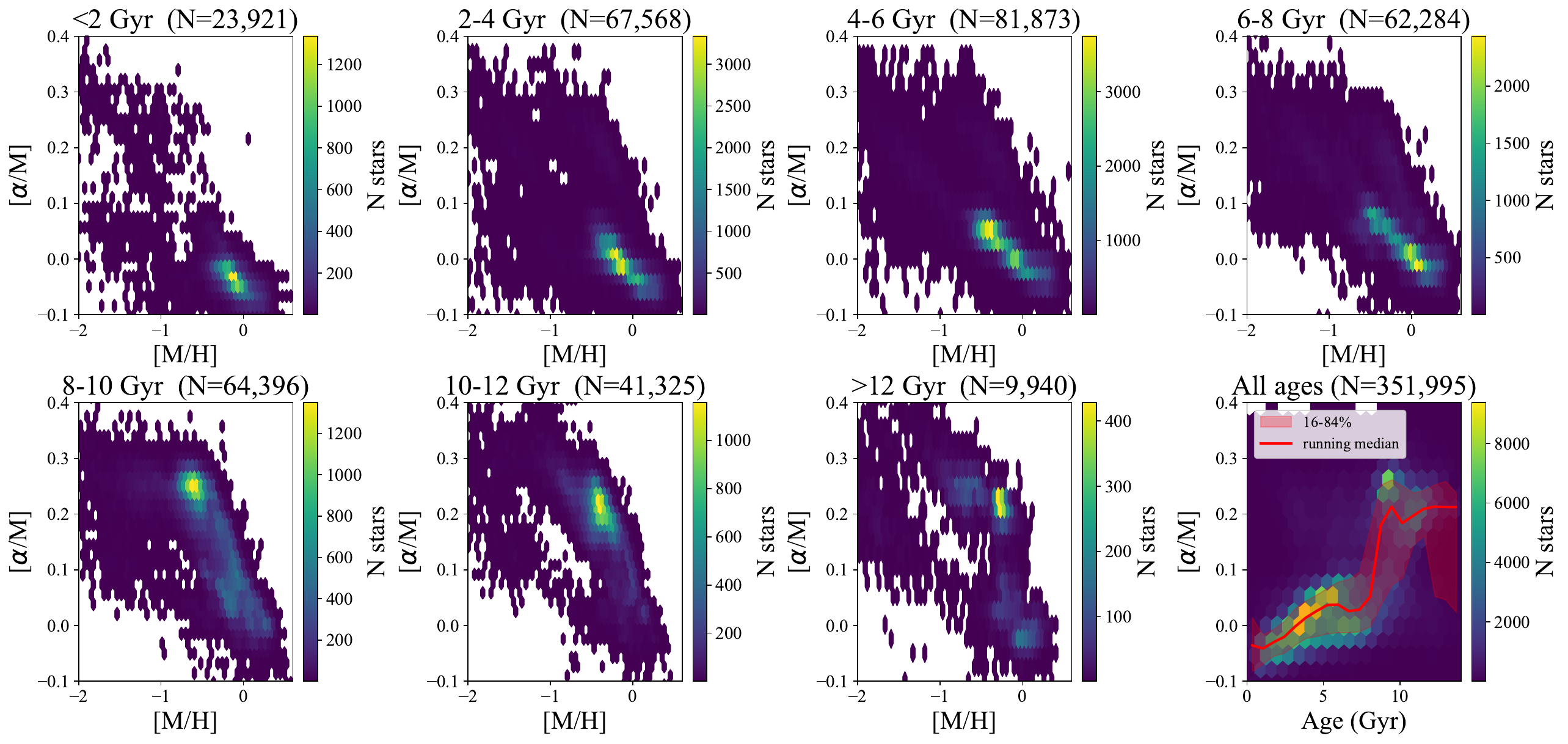}
    \caption{[$\alpha$/M] versus [M/H] plots for each age bin in our sample from the youngest (upper left) to the oldest (lower row, second from the right) showing the evolution of chemistry over time in the galaxy, with the number of stars at each composition indicated by color. In the lower right panel, we show the evolution of the $\alpha$-element enhancement with age in the sample, with a running median (solid line) and uncertainty band (16-84th percentile) shown in red.} 
    \label{fig:agemapalpha}
\end{figure*}

\begin{figure*}[tbh]
    \centering  \includegraphics[width=0.99\textwidth,clip=true, trim=0in 0in 0in 0in]{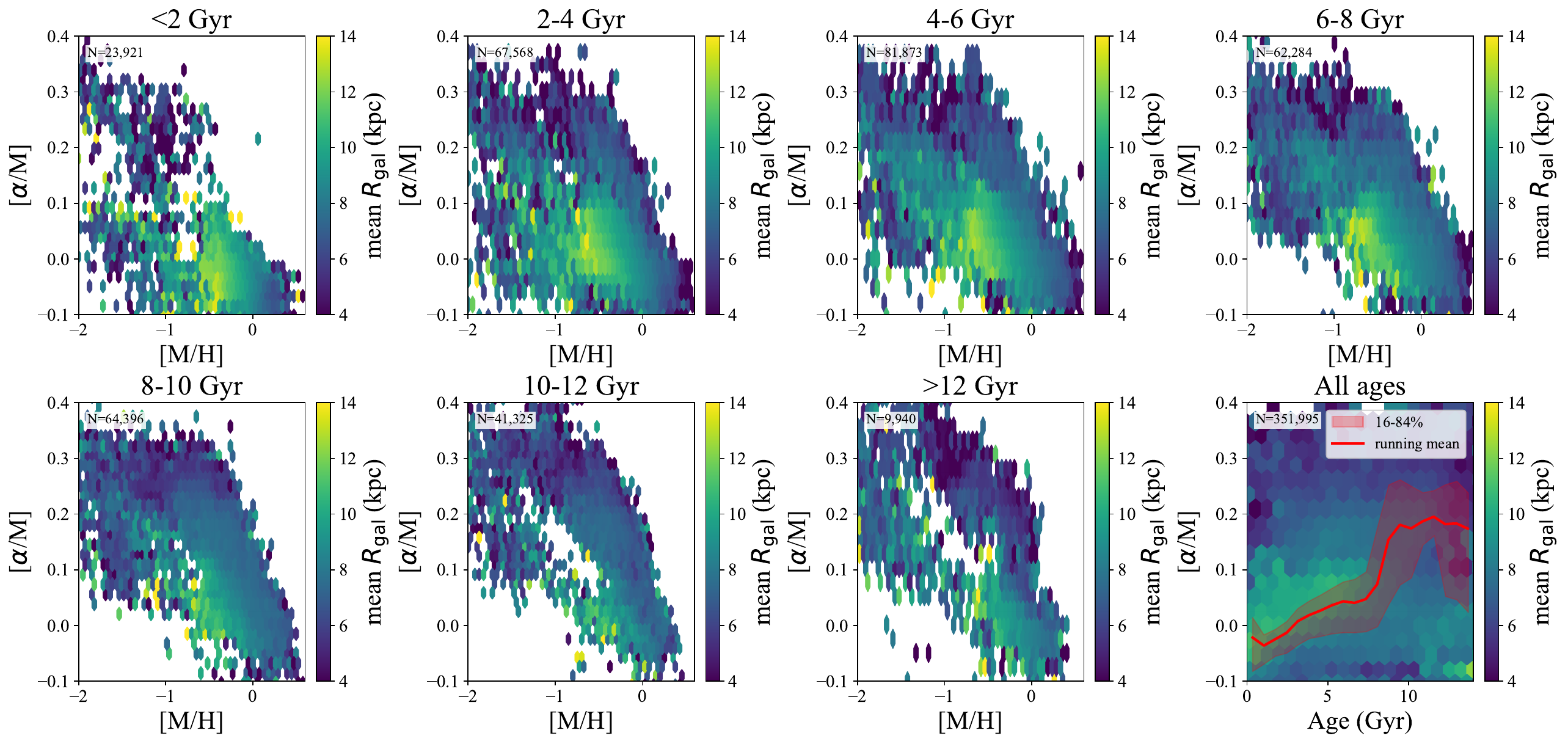}
    \caption{Similar to Figure \ref{fig:agemapalpha}, [$\alpha$/M] versus [M/H] plots for each age bin in our sample from the youngest (upper left) to the oldest (lower row, second from the right) showing the evolution of chemistry over time in the galaxy, but in this case the average galactocentric radius of each bin indicated by color. In the lower right panel, we show the evolution of the $\alpha$-element enhancement with age in the sample, with a running median (solid line) and uncertainty band (16-84th percentile) shown in red.} 
    \label{fig:agemapalphaRgal}
\end{figure*}

\begin{figure*}[tbh]
    \centering  \includegraphics[width=0.99\textwidth,clip=true, trim=0in 0in 0in 0in]{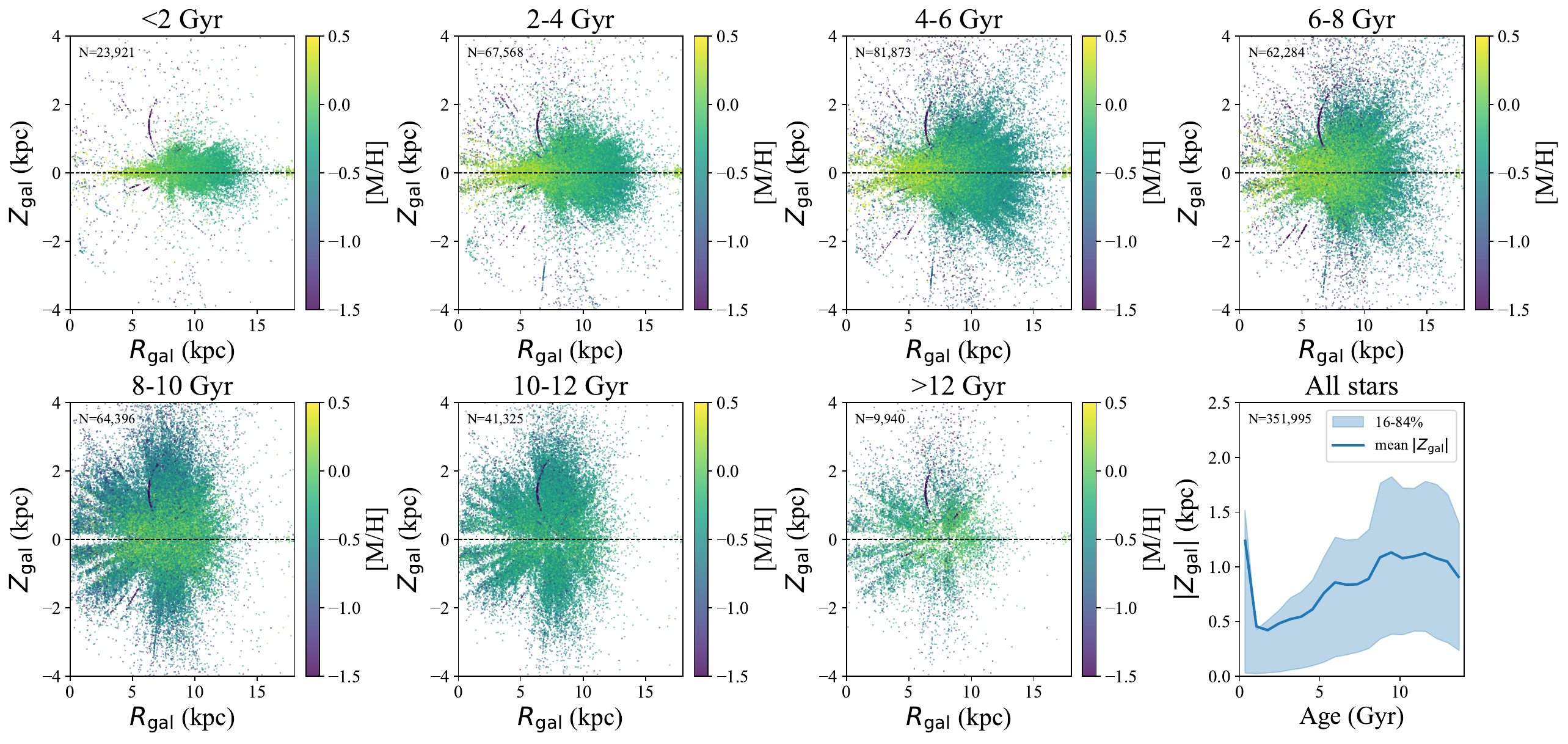}
    \caption{Galactic height versus galactocentric radius plots for our stars, broken up into bins of age from the youngest (upper left) to the oldest (lower panel, second from right). In each panel, stars are color coded by their metallicity, indicating the evolution of chemistry with position and time. In the lower right panel, we show the evolution of the average height above the plane in our sample with age, with a running median and uncertainty band (16-84th percentile) shown in blue.}
    \label{fig:agemapz}
\end{figure*}

In Figure \ref{fig:agemapkiel}, we show the Kiel diagrams for subsamples at various metallicities, color-coded by age. In each of these metallicity bins, we see a coherent pattern, with ages of the first ascent red giant branch correlating with temperature in the expected way. The core-helium-burning stars follow a different pattern as predicted, with the youngest core-helium-burning stars near solar metallicity populating the secondary clump \citep{Girardi1999}, and older core helium burning stars at lower gravity. These patterns are extremely consistent with theoretical expectations \citep{Pinsonneault2026} and well-vetted datasets \citep{Pinsonneault2025}, adding additional confidence in our inferred ages.

In Figure \ref{fig:agemapalpha}, we show the chemical makeup of the galaxy at different times. At the oldest ages (Age $>8$ Gyr, see lower panels), the galaxy was dominated by $\alpha$-element enhanced stars, which received significant enrichment from Type II supernovae, but limited contributions from Type Ia supernovae. This is generally consistent with our current picture of galactic chemical evolution \citep[e.g.][]{Weinberg2019}. However, even at the oldest age, we see a population of relatively metal-rich and $\alpha$-rich stars. At this time, we suspect that these stars represent a combination of sources from 1) binary interactions \citep{LiY2022, Bufanda2023, Frazer2025}, where a companion has stripped away mass, making the star appear older than it is and 2) truly old stars from the bulge or inner galaxy, where chemical enrichment proceeded rapidly \citep{Griffith2021, Joyce2023}, that have migrated radially in the intervening time. In Figure \ref{fig:agemapalphaRgal} we color each bin by the average galactocentric radius, and note that these $\alpha$-poor, metal-rich old stars tend to be in the solar neighborhood rather than towards the inner galaxy, suggesting that binary interactions may be the dominant source of such stars in our sample.

We can also look at the current positions of the stars as a function of age and chemistry (Figure \ref{fig:agemapz}). We find that the oldest stars in the galaxy tend to be at larger $|$Z$|$ heights on average, consistent with either the upside-down formation theory of galaxies \citep{Bird2013} or dynamical heating over time \citep{Mackereth2019}. The higher heights above the plane, many of which are populated by thick-disk and halo stars, become much rarer for stars younger than $\sim$ 8 Gyr, and the younger thin disk stars are also generally $\alpha$-poor. When looking at the youngest stars ($<$2 Gyr), which have generally undergone significantly less radial migration, the radial metallicity gradient is strongly apparent, with star formation happening at higher metallicity in the inner galaxy and lower metallicity in the outer galaxy. These overall patterns are consistent with previous work, but increases in the sample size, resolution, precision, and accuracy of such maps offers an opportunity for more detailed study of the galaxy's evolution.


\section{Conclusions} 
In this work, we have evaluated the impact of machine learning methodology on the ages inferred for red giant stars from spectroscopic parameters. We have found that:
\begin{itemize}
    \item Even simple machine learning architectures are sufficient to infer reasonable ages for these stars (offsets $<0.5$ Gyr, scatter $\sim$ 2 Gyr).
    \item The exact hyperparameters chosen for the network are not very important (offsets $<0.3$ Gyr, scatter $<1.5$ Gyr).
    \item The training set chosen matters somewhat (offset usually $<0.5$ Gyr, but can be larger).
    \item These ages are generally accurate for ages less than 9 Gyr (offset $\sim$ 0.5 Gyr, scatter $\sim1$ Gyr).
    \item At older ages, the results are less consistent with seismic and cluster measurements, and have larger method-to-method systematics (offsets $\sim$ several Gyr, scatter $\sim$ several Gyr). 
    \item Results are also less reliable at low temperatures, low metallicities, low gravities, and in the innermost and outermost portions of the galaxy.
    \item Where ages are reliable, we are able to make coherent maps of the galaxy's evolution, with our data showing the expected trends in $\alpha$-element abundance, scale height, and chemical gradients.
\end{itemize}

In the future, we hope that large datasets like this, with kinematic, chemical, and age information, will help facilitate quantitative comparisons to chemical evolution models \citep{JohnsonJ2025a}, improve our understanding of dynamical heating and radial migration \citep{Lian2022}, and allow detailed comparison to simulations of the evolution of Milky Way-like galaxies \citep{Sanderson2020, Beaton2022} in order to solidify our understanding of how the Milky Way evolved. 

We also hope that ongoing work in a variety of domains will continue improving our ability to precisely and accurately estimate ages for large samples of giant stars. On the stellar modeling side, efforts to compare models \citep{SilvaAguirre2020, Morales2025}, understand their uncertainties \citep{Ying2023, LiY2024}, and compare them to benchmark systems \citep{Tayar2017, Joyce2018b, Schimak2026} can continue to improve our estimated ages. Seismic results from current \citep[e.g. TESS,][]{Ricker2015}, upcoming \citep[e.g. PLATO and Roman,][]{Rauer2014, Roman} and proposed \citep[e.g. HAYDN, ][]{HAYDN} space-based missions can continue to increase our training samples. Efforts to expand the seismology that can be done from the ground are also ongoing \citep{Grundahl2006, Auge2020, Hey2025, LiY2025b,Schochet2026}, and it may be possible to use large datasets such as Rubin LSST \citep{RubinLSST} to expand such work. Additional results will soon be released by Gaia (Data Release 4), which will improve seismic calibration \citep{Zinn2019, Pinsonneault2025} and can be combined with seismic results to infer ages for larger samples \citep{Theodoridis2025}. Finally, current \citep[e.g. LAMOST, GALAH, SDSS V,][]{Cui2012LAMOST, GalahDR4, Kollmeier2026} and upcoming  \citep[e.g. 4MOST, WEAVE,][]{4MOST,WEAVE} surveys continue to release ever more spectra, promising to increase the sample for galactic mapping investigations to millions of stars. This will allow more detailed division of the galaxy, as well as probing regions, such as the bulge, halo, and far side of the disk, that have been poorly studied in previous work. Long term, we expect these age samples to contribute to better understanding of our own Cosmic Origins and to facilitate the use of our Milky Way as a model to understand galactic formation and evolution more generally, from our own Local Group out to the distant galaxies at the beginning of the Universe.


\section*{Acknowledgments}
We thank D. Acosta, N. Camargo, A. Camazon, S. Candiani, S. Carlton-Jones, R. Faeza,  J. Gregory, Z. Knutson,   N. Meyer, J. Morley, A. Mujumdar, J. Rivero-Aleman, S. Santos, M. Shapiro, R. Souchet, C. Wagner, and Z. Yates for contributing to initial discussions about this work.

We thank D. Horta for helpful suggestions about network interpretation.

Funding for the Sloan Digital Sky Survey V has been provided by the Alfred P. Sloan Foundation, the Heising-Simons Foundation, the National Science Foundation, and the Participating Institutions. SDSS acknowledges support and resources from the Center for High-Performance Computing at the University of Utah. SDSS telescopes are located at Apache Point Observatory, funded by the Astrophysical Research Consortium and operated by New Mexico State University, and at Las Campanas Observatory, operated by the Carnegie Institution for Science. The SDSS web site is \url{www.sdss.org}.

SDSS is managed by the Astrophysical Research Consortium for the Participating Institutions of the SDSS Collaboration, including the Carnegie Institution for Science, Chilean National Time Allocation Committee (CNTAC) ratified researchers, Caltech, the Gotham Participation Group, Harvard University, Heidelberg University, The Flatiron Institute, The Johns Hopkins University, L'Ecole polytechnique f\'{e}d\'{e}rale de Lausanne (EPFL), Leibniz-Institut f\"{u}r Astrophysik Potsdam (AIP), Max-Planck-Institut f\"{u}r Astronomie (MPIA Heidelberg), Max-Planck-Institut f\"{u}r Extraterrestrische Physik (MPE), Nanjing University, National Astronomical Observatories of China (NAOC), New Mexico State University, The Ohio State University, Pennsylvania State University, Smithsonian Astrophysical Observatory, Space Telescope Science Institute (STScI), the Stellar Astrophysics Participation Group, Universidad Nacional Aut\'{o}noma de M\'{e}xico, University of Arizona, University of Colorado Boulder, University of Illinois at Urbana-Champaign, University of Toronto, University of Utah, University of Virginia, Yale University, and Yunnan University.

\section*{Author Contributions}

J. Tayar conceived the project, taught the AST 4300 Galactic Astronomy course where the initial investigations of this idea were done, provided overall supervision for the analysis, and wrote the majority of the text. C. Mankowski created Figures 7,8 and 11-16.
L. Tunca revised Figures 1 and 2, created Figures 3 through 6 and 9, along with Table 1, investigated issues with the APO-K2 data set, wrote drafts for several subsections, selected the tested network parameters, trained the machine learning models, and assessed which model performed best. D. Jordan made Figure 2. 
M. Severino did preliminary development and standardization of the paper’s visual presentation, including figure styling, color schemes, and consistency across graphical elements. 
A. Leddy contributed to the figure design.
S. McArthur, Z.Benton, and S. Armstrong contributed to the cluster results. 
S. McArthur and E. Bower contributed to the decisions of how to treat binarity.

D. Jordan, Z.Benton, A. Leddy Z. Sanchez, C. Avery, E. Cummings, J. Donley, R. Freeman, D. Fulcher, V. Hervie, J. Ivey, H. Kenney, W. MacMillan, J. Mahoney, E. Maramba, E. Philip, Y. Simpson, and E. Strojie contributed to the initial explorations of the variations between networks, training sets, and/or the behavior of specific subsamples.

All authors have read and agree with the final paper draft.

\section*{Software and third party data repository citations} \label{sec:cite}

This research made use of \texttt{TensorFlow} \citep{tensorflow2015-whitepaper},  \texttt{astropy} \citep{astropy2013,astropy2018,astropy2022}
and visualization and analysis packages including 
\texttt{numpy} \citep{numpy} and \texttt{matplotlib} \citep{Matplotlib}.

\bibliography{Tayar2026.bbl}{}
\bibliographystyle{aasjournalv7}

\end{document}